\pdfoutput=1

\documentclass[11pt,twoside,a4paper,cmspaper,final,collab]{cms-tdr}
\def\svnVersion{461800P}\def\cmsCernNoTag{CERN-EP-2017-330}\def\cmsCernDate{\today}\def\cmsMessage{Published in the Journal of High Energy Physics as \href{http://dx.doi.org/10.1007/JHEP05(2018)127}{\doi{10.1007/JHEP05(2018)127.}}}

\begin{document}\cmsNoteHeader{EXO-16-004}

\hyphenation{had-ron-i-za-tion}
\hyphenation{cal-or-i-me-ter}
\hyphenation{de-vices}
\RCS$HeadURL: svn+ssh://svn.cern.ch/reps/tdr2/papers/EXO-16-004/trunk/EXO-16-004.tex $
\RCS$Id: EXO-16-004.tex 456277 2018-04-19 10:08:39Z jalimena $
\newlength\cmsFigWidth
\ifthenelse{\boolean{cms@external}}{\setlength\cmsFigWidth{0.85\columnwidth}}{\setlength\cmsFigWidth{0.4\textwidth}}
\ifthenelse{\boolean{cms@external}}{\providecommand{\cmsLeft}{top\xspace}}{\providecommand{\cmsLeft}{left\xspace}}
\ifthenelse{\boolean{cms@external}}{\providecommand{\cmsRight}{bottom\xspace}}{\providecommand{\cmsRight}{right\xspace}}
\cmsNoteHeader{EXO-16-004} \def\neutralino  {\ensuremath{\tilde{\chi}^0}\xspace}
\def\gluino      {\ensuremath{\tilde{g}}\xspace}
\def\stop        {\ensuremath{\tilde{t}}\xspace}

\title{Search for decays of stopped exotic long-lived particles produced in proton-proton collisions at $\sqrt{s}=13\TeV$}

\date{\today}

\abstract{
A search is presented for the decays of heavy exotic long-lived particles (LLPs) that are produced in proton-proton collisions at a center-of-mass energy of 13\TeV at the CERN LHC and come to rest in the CMS detector. Their decays would be visible during periods of time well separated from proton-proton collisions. Two decay scenarios of stopped LLPs are explored: a hadronic decay detected in the calorimeter and a decay into muons detected in the muon system. The calorimeter (muon) search covers a period of sensitivity totaling 721\,(744) hours in 38.6\,(39.0)\fbinv of data collected by the CMS detector in 2015 and 2016. The results are interpreted in several scenarios that predict LLPs. Production cross section limits are set as a function of the mean proper lifetime and the mass of the LLPs, for lifetimes between 100\unit{ns} and 10 days. These are the most stringent limits to date on the mass of hadronically decaying stopped LLPs, and this is the first search at the LHC for stopped LLPs that decay to muons.
}

\hypersetup{%
pdfauthor={CMS Collaboration},%
pdftitle={Search for decays of stopped exotic long-lived particles produced in proton-proton collisions at sqrt(s)=13 TeV},%
pdfsubject={CMS},%
pdfkeywords={CMS, physics, stopped particles}}

\maketitle
\section{Introduction}
Heavy long-lived particles (LLPs) on the order of 100\GeV are not present in the standard model (SM).
Therefore, any sign of them would be an indication
of new physics.
Many extensions of the SM predict the existence of LLPs~\cite{Dimopoulos:1996vz,Baer:1998pg,Jittoh:2005pq,Fairbairn:2006gg,Strassler:2006im,Arvanitaki:2008hq,ArkaniHamed:2004fb,Arvanitaki:2012ps}.
At the CERN LHC, the LLPs will stop inside the detector material if they lose all of their kinetic energy
while traversing the detector,
which will typically occur for particles with initial
velocities less than about $0.5c$~\cite{Arvanitaki:2005nq}.
This energy loss can occur via nuclear interactions if they are strongly interacting and/or through ionization if they are charged.
The observation of a stopped particle decay signature would not only indicate new physics but also help measure the lifetime of LLPs, giving insights into various beyond the standard model (BSM) theories.

If these stopped LLPs have lifetimes longer than tens of nanoseconds, most of their decays would be reconstructed as separate events unrelated to their production~\cite{Graham_stoppedParticlesAtLHC}. Owing to the difficulty of differentiating between the LLP decay products and SM particles from LHC proton-proton ($\Pp\Pp$) collisions, these subsequent decays are most easily identified when there are no proton bunches in the detector. The detector is quiet during these out-of-collision time periods with the exception of rare noncollision backgrounds, such as cosmic rays, beam halo particles, and detector noise. If LLPs come to a stop in the detector, they are most likely to do so in the densest detector materials, which in the CMS detector are the electromagnetic calorimeter (ECAL), the hadron calorimeter (HCAL), and the steel yoke in the muon system. If the stopped LLPs decay in the calorimeters, relatively large energy deposits occurring in the intervals between collisions could be observed. Furthermore, if the stopped LLPs decay into muons, displaced muon tracks out of time with the collisions could be detected.

In this paper we present two searches for stopped LLPs that decay out of time with respect to the presence of proton bunches in the detector. One search targets hadronic decays detected in the calorimeters, and the other looks for decays to muon pairs in the muon system. These two search channels are analyzed independently using data collected by the CMS experiment in 2015 and 2016 with separate dedicated triggers. The calorimeter (muon) search uses $\sqrt{s}=13\TeV$ data corresponding to an integrated luminosity of 38.6\,(39.0)\fbinv collected with LHC $\Pp\Pp$ collisions separated by 25\unit{ns} during a search interval totaling 721\,(744) hours. The size of the search sample is further reduced by applying a series of offline selection criteria to decrease the number of events that most likely come from the primary sources of background.

The calorimeter search presented here improves upon previous searches performed by the CMS Collaboration, the most recent of which used $\sqrt{s}=8\TeV$ $\Pp\Pp$ collision data corresponding to an integrated luminosity of 18.6\fbinv collected in 2012~\cite{StoppedParticlesCMS8TeV}. This search excluded long-lived gluinos ($\PSg$) with masses below 880\GeV and long-lived top squarks ($\PSQt$) with masses below 470\GeV, for lifetimes between $10\mus$ and 1000\unit{s}. The results of earlier, similar searches have been reported by the D0 Collaboration at the Tevatron~\cite{stoppedGluinos_D0_2011} and by the CMS~\cite{CMSstoppedParticles7TeV,Khachatryan:2010uf} and ATLAS Collaborations~\cite{stoppedParticles_ATLAS_2012,stoppedParticles_ATLAS_2013}. The displaced muon search is newly added to investigate different models with leptonic decays of stopped LLPs, such as those of gluinos~\cite{Arvanitaki:2005nq} and multiply charged massive particles (MCHAMPs)~\cite{0264-9381-23-24-008,DERUJULA1990173,PhysRevD.41.2388,Langacker:2011db}. Searches for decays of stopped LLPs are complementary to searches for heavy stable charged particles (HSCPs) that pass through the detector and can be identified by their energy loss and time-of-flight (TOF) information~\cite{cmllp_ATLAS_PLB_heavylonglived_2011,cmllp_ATLAS_PLB_hadronizingSquarksGluinos_2011,cmllp_ATLAS_PLB_highlyionizing_2011,cmllp_ATLAS_sleptonsRhadrons_2012,cmllp_ATLAS_multiplyCharged_2013,cmllp_ATLAS_multiplyCharged_2015,cmllp_ATLAS_EPJC_heavylonglived_2015,cmllp_CMS_JHEP_2013,cmllp_CMS_JHEP_2011,cmllp_CMS_fractionallyCharged_2013,cmllp_CMS_PLB_2012,cmllp_CMS_reinterpretation_EPJC_2015,cmllp_LHCb_2015,PhysRevD.94.112004}. The searches presented here would allow the study of the decay of such heavy particles, whereas dedicated HSCP searches typically look for the particle itself, before it decays. However, both the searches for decays of stopped LLPs and for HSCPs are sensitive to a similar range of lifetimes.
\section{The CMS detector}
\label{sec:detector}
The central feature of the CMS apparatus is a superconducting solenoid of
6\unit{m} internal diameter, providing a magnetic field of 3.8\unit{T}.
Within the solenoid volume are a silicon pixel and strip
tracker, a lead tungstate crystal ECAL, and
a brass and scintillator HCAL, each composed of a barrel
and two endcap sections. Forward calorimeters extend the pseudorapidity $\eta$ coverage
provided by the barrel and endcap detectors. In the region $\abs{\eta} < 1.74$,
the HCAL cells have widths of 0.087 in $\eta$ and 0.087 radians in azimuth ($\phi$).
In the $\eta$-$\phi$ plane, and for $\abs{\eta} < 1.48$, the HCAL cells map on to
$5 \times 5$ arrays of ECAL crystals to form calorimeter towers projecting radially
outwards from close to the nominal $\Pp\Pp$ collision interaction point (IP). For $\abs{\eta} > 1.74$, the
coverage of the towers increases progressively to a maximum of 0.174 in $\Delta \eta$
and $\Delta \phi$. Within each tower, the energy deposits in ECAL and HCAL cells are
summed to define the calorimeter tower energies, which are subsequently used to provide the
energies and directions of hadronic jets. In the HCAL barrel (HB) and endcap, scintillation light is detected
by hybrid photodiodes (HPDs), and each HPD collects signals from 18 different HCAL channels.
Signals from four HPDs are then digitized by analog-to-digital converters within a single readout box (RBX).

Muons are measured in gas-ionization
chambers embedded in the steel flux-return yoke outside the solenoid.
Muons are measured in the range $\abs{\eta}< 2.4$, with detection
planes made using three technologies: drift tubes (DTs) in the barrel, cathode strip chambers (CSCs) in the endcaps, and
resistive plate chambers (RPCs) in both the barrel and the endcaps. All these technologies provide both position and
timing information.
Hits within each DT or CSC chamber are matched to form a reconstructed DT or CSC segment.

The first level (L1) of the CMS trigger system, composed of custom hardware processors,
uses information from the calorimeters and muon detectors to select the most interesting
events in a fixed time interval of less than 4\mus. The high-level trigger
processor farm further decreases the event rate from around 100\unit{kHz} to less than 1\unit{kHz}, before data storage.

A more detailed description of the CMS detector, together with a definition of the
coordinate system used and the relevant kinematic variables, can be found in Ref.~\cite{Chatrchyan:2008zzk}.
\section{Data and Monte Carlo simulation}
\label{sec:montecarlo}

\subsection{Data samples}
The LHC accelerates two proton beams in opposite directions such that the protons collide at several
points along the LHC ring, including one at the CMS detector.
Each LHC beam consists of a number of proton bunches arranged into an irregular pattern of ``trains''~\cite{Bailey:691782}.
Within a train, the proton bunches are nominally spaced 25\unit{ns} apart, with a larger spacing between trains to account for the needs of the injection process.
In an LHC orbit there are 3564 bunch slots (BXs), which are 25\unit{ns} long.
Each BX could be filled with proton bunches, which usually occupy the first 2.5\unit{ns} of the BX, or could be empty.
The trains may be spaced such that there could be multiple empty BXs between filled BXs.
To search for LLP decays during
these empty BXs, dedicated triggers select events at least two BXs
away from any proton bunches.
Thus these triggers are live only during these specific time windows.
This distance of two BXs is chosen so that we maximize the search
time window while suppressing most of the events from secondary pp interactions and from ``beam halo'',
which are mostly muons
traveling outside the LHC beam that are produced by LHC beam--collimator
scattering.

The search is performed with $\sqrt{s}=13\TeV$ $\Pp\Pp$ collision run data collected by the CMS experiment in 2015 and 2016.
The 2015 calorimeter (muon) search sample, taken between August and November 2015, corresponds to an integrated luminosity of
2.7\,(2.8)\fbinv and spans a trigger livetime, which is the amount of time the triggers are
live in between collisions, of 135\,(155) hours. The 2016 calorimeter (muon)
search sample was taken between May and October 2016, during which a data sample corresponding to an integrated luminosity of 35.9\,(36.2)\fbinv was
recorded, spanning a trigger livetime of 586 (589) hours. We do not consider the possibility of LLPs that were produced in 2015 but decayed in 2016. In both the 2015 and 2016 searches, we use cosmic run data
collected by dedicated triggers as a control sample.
These dedicated cosmic run data were recorded during LHC machine technical stops,
several days after collision runs. A negligible amount of long-lived signal produced during collisions could
have decayed during these cosmic runs for the lifetimes considered in this analysis.
The instrumental noise background estimate is extrapolated from the instrumental noise measured in these control samples.
Most of the other sources of background are estimated from sideband regions of the main data sample,
except for the cosmic ray muon background in the calorimeter search, which is estimated from MC simulation.

\subsection{Benchmark models}
Several simplified models are considered in this search, and samples are generated for each using Monte Carlo (MC) simulation.

In the calorimeter search, we interpret the results in the context of two-body ($\PSg \to \cPg\PSGcz$) and three-body ($\PSg \to \cPq\cPaq\PSGcz$) decays of a gluino into the lightest supersymmetric (SUSY) particle (LSP), the neutralino ($\PSGcz$). Long-lived gluinos are predicted by ``split SUSY''~\cite{Giudice_SplitSUSY,ArkaniHamed:2004yi}, in which gauginos have relatively small masses with respect to sfermions, which could be massive, since SUSY is broken at a scale much higher than the weak scale. This large mass splitting causes the long lifetime of the gluinos, since gluinos can only decay via a virtual squark. We also consider the decay of a long-lived top squark ($\PSQt \rightarrow \cPqt\PSGcz$) that can be the next-to-LSP particle (NLSP) in various dark matter scenarios~\cite{Abercrombie:2015wmb,Carpenter:2016thc,Covi:2014fba}. Here the LSP should be loosely interpreted as any new, neutral, non-interacting fermion, and not necessarily as a SUSY neutralino.

In the muon search, we consider a different model for a three-body decay of the gluino ($\PSg \to \cPq\cPaq\PSGczDt, \PSGczDt \to \MM\PSGcz$), which is complementary to the calorimeter search.
In this model, the mass of the LSP neutralino ($\PSGcz$) is chosen to be 0.25 times the gluino mass, and the mass of the
NLSP neutralino ($\PSGczDt$) is chosen to be 2.5 times the LSP neutralino mass.
A second simplified model used in the muon search predicts exotic particles called MCHAMPs, whose charges are multiples of the elementary charge $e$ and which are predicted by several BSM theories~\cite{Langacker:2011db}. We assume an MCHAMP with charge $\abs{Q}=2e$ decays into two same-sign muons ($\text{MCHAMP} \to \mu^{\pm}\mu^{\pm}$).

\subsection{Signal generation}

The signal generation process is divided into three major stages.
In Stage 1, the LLPs for each signal process are generated from $\Pp\Pp$ collisions
with \PYTHIA~\cite{pythia6,pythia8} and propagated through
the detector with \GEANTfour v9.2~\cite{geant4_simulation_toolkit,geant4_developments}.
For the MCHAMP signal, \PYTHIA v6.4 is used, while for the gluino and top squark signals, \PYTHIA v8.205 is used.
If the LLPs are strongly interacting, as in the case of the gluinos and top squarks, they hadronize into R-hadrons~\cite{Fayet_susy,Fayet_gluinos,Farrar_phenomenology_hadronicStates_susy} upon production, whose interaction with the CMS detector in the simulation is described by the cloud model~\cite{Kraan:2004tz,Mackeprang:2006gx}. In this model, R-hadrons are treated as SUSY particles surrounded by a cloud of loosely bound quarks and gluons. The fraction of produced R-hadrons that contain a gluino and a valence gluon is set to 10\%, a convention used in previous analyses~\cite{StoppedParticlesCMS8TeV,cmllp_ATLAS_PLB_heavylonglived_2011}. However, because the R-hadrons interact an average of ten times in the calorimeter, their flavor is effectively randomized. 
Some fraction of these R-hadrons are sufficiently slow moving to come to a
stop in the detector material.
Because they are doubly charged, MCHAMPs
ionize heavily and thus a significant number also stop in the
detector.

In Stage 2, the parent LLP
or R-hadron is constrained to decay at the stopping position defined
in Stage 1. The LLP decay is simulated by a second \GEANTfour step,
and the decay products are propagated through the detector.

Finally, in Stage 3, a pseudo-experiment
MC simulation is conducted to estimate the probability for stopped particle
decays to occur in the time window between collisions when data is being collected.
The Stage 3 MC simulation determines an effective integrated luminosity by using the good data-taking periods
and the LHC filling scheme to calculate the fraction of stopped particle decays that
occur when the trigger is live. For a given particle
lifetime, the effective integrated luminosity
is defined as the total integrated luminosity multiplied by the probability
that the particle decays at a time when the trigger is live in between collisions.
In other words,
Stages 1 and 2 determine how the signal will look in the detector,
and Stage 3 determines when it will occur. More details on the signal generation process are given in Refs.~\cite{StoppedParticlesCMS8TeV,CMSstoppedParticles7TeV,Khachatryan:2010uf}.

\section{Event selection}
The calorimeter search and the muon search employ different search strategies and thus different selection criteria,
which are described in turn below.

\subsection{Calorimeter search}
In the calorimeter search, we look for hadronic decays of LLPs in the calorimeter that produce
energy deposits that could be reconstructed as at least one high-energy jet.
We trigger on calorimeter jets with energy greater than 50\GeV and $\abs{\eta} < 3$ that are at least two BXs away from $\Pp\Pp$ collisions.

The major background sources are cosmic rays, beam halo, and HCAL noise.
Cosmic ray and beam halo muons can emit a shower of photons via bremsstrahlung, which could be reconstructed as a jet and mistaken for signal.
HCAL noise~\cite{2010JInst...5T3014C} can give rise to spurious signals, which in the barrel could appear in one or several HPDs within a single RBX,
and thus be incorrectly reconstructed as a jet. We observe that the rate of each of these background sources drops exponentially as a function of the jet energy.
We thus require the events to have a leading (highest energy) calorimeter-based jet with energy greater than 70\GeV.
The calorimeter-based jets are reconstructed using an anti-\kt clustering algorithm~\cite{Cacciari:2008gp,Cacciari2012}
with a distance parameter of 0.4. To increase the sensitivity of the search, we require that the leading jet in each event
is located within $\abs{\eta} < 1.0$, where R-hadrons are more likely to stop and where there is relatively less background from beam halo.

Secondary background sources include out-of-time collisions from remnant protons between bunches, and beam-gas interactions in the detector. The rate of these secondary background events becomes negligible after we require that there are no reconstructed collision vertices in the events.

Cosmic ray muon events usually feature a large number of reconstructed DT segments and RPC hits, whereas signal events in the calorimeter search would not. We exploit this difference to distinguish signal events from cosmic ray muons. While it is possible for the hadronic shower of an R-hadron decay to pass through the first layers of the iron yoke and induce reconstructed DT segments, these DT segments are located only in the inner layers of the muon chambers ($r < 560\cm$, where $r$ is the transverse distance to the IP) and cluster near the leading jet. On the other hand, cosmic ray muons are equally likely to leave DT segments in all layers in both the upper and lower hemispheres of the muon system, and the angle between the jet and DT segments in $\phi$ is more evenly distributed. As a result, we are able to substantially reduce the cosmic ray muon background contamination in the signal region by rejecting events that have at least two DT segments in the outermost barrel layer of the muon system, events that have any DT segments in the second outermost barrel layer, events that have two DT segments with a large separation in $\phi$ ($\abs{\Delta\phi} > \pi/2$), events that have DT segments in the three innermost layers that are separated in $\phi$ from the leading jet by at least 1.0 radian, and events that have close-by RPC hits in different layers ($\DR = \sqrt{\smash[b]{(\Delta\phi)^2 + (\Delta\eta)^2}} < 0.2$  and  $\Delta r > 0.5\unit{m}$). We make looser DT segment requirements in the outermost than in the second outermost layer because signals are very likely to coincide with standalone DT segments that are not from cosmic ray muons but particles from the $\Pp\Pp$ collision. Most of these standalone DT segments from the $\Pp\Pp$ collision are located in the outermost muon barrel layer. With these selection criteria, we are able to avoid incorrectly rejecting signal events, thus increasing the signal efficiency, while still rejecting most of the cosmic ray muon events.

Beam halo muons
travel closely along the beam pipe, typically traversing both sides of the muon endcap systems and resulting in a few reconstructed CSC segments. Therefore, we veto events with any CSC segments having at least five reconstructed hits. As will be discussed in Section~\ref{sec:signalEfficiency}, since signal events may include some CSC segments, requiring a minimum number of CSC hits in the veto avoids a loss of signal efficiency.

Random electronic noise in the HCAL gives rise to events in which the time response of the HCAL readout is very different from the well-defined response from particles showering in the calorimeter. This HCAL noise creates spurious clustered energy deposits that can be reconstructed as a jet, which would contaminate the signal region and therefore should be removed. Analog signal pulses produced by the HCAL electronics are read out over ten BXs centered around the pulse maximum. The pulse shape from showering particles consists of a peak at the collision BX and an exponential decay over the subsequent BXs. Particle showers create clustered energy deposits spread over several neighboring calorimeter towers in $z$ and $\phi$, while noise produces deposits in just one or two towers, or several towers in a single HPD or RBX. In addition to the standard HCAL noise filter~\cite{2010JInst...5T3014C}, we use a series of offline selection criteria that exploit these timing and topological characteristics to remove the HCAL noise events. These criteria are described in detail in Ref.~\cite{Khachatryan:2010uf}.

\subsection{Muon search}
In the muon search, we look for LLPs where the decay products include two muons. We expect the signal
to look like a pair of muons originating anywhere in the detector material, but displaced from the IP.
The muons would be back-to-back in the two-body MCHAMP decay, but not for the three-body gluino decay.

The primary background sources in the muon search include cosmic ray muons, beam halo, and muon detector noise. The latter two background sources are negligible after we apply the full selection.

The trigger used in the muon search selects events at least two BXs away
from the $\Pp\Pp$ collision time with at least one muon reconstructed in the muon system,
whose transverse momentum $\pt$ is at least 40\GeV.
As in the calorimeter search, we select events offline that have no reconstructed collision vertices.

Tracks that are reconstructed using only hits in the muon system are called standalone muon tracks~\cite{muonReco7TeV}.
However, the standard standalone track reconstruction assumes that muons originate from the IP,
which is inappropriate for displaced muon searches.
As a result, a new muon reconstruction algorithm was
developed for this analysis, which produces displaced standalone (DSA) muon tracks~\cite{cms_muonReco_Run2_FirstResults2015Data_EPS2015}.
The DSA tracks are reconstructed using only hits in the muon detector, and they have no constraints to the IP.
Thus, DSA tracks are truly using only the muon system.

We require events to have exactly one good DSA track in the upper hemisphere of the detector and exactly one good DSA track in the lower hemisphere. Both DSA tracks must have $\pt>50\GeV$, at least three DT chambers with valid hits, and at least three valid RPC hits. To reduce the background from beam halo, the DSA tracks must also have zero valid CSC hits.

Timing information in the DTs and RPCs, indicating whether the muon
is incoming toward the detector center or outgoing away from the detector center, is used to distinguish muons from a signal event from the
cosmic ray muon background.
Cosmic ray muons are predominantly incoming when traversing the upper
hemisphere and outgoing when traversing the lower hemisphere, as they come in from above the detector
and continue to move downwards. Muons from a signal event, on the
other hand, would be outgoing in both hemispheres.

We place selection criteria on both the upper and lower hemisphere DSA tracks in order to obtain a
good time measurement.
We require at least eight independent time measurements for the TOF computation.
We require that the uncertainty in the time measured at the IP for DSA tracks, assuming the muon is outgoing,
is less than 5.0\unit{ns}.

Next, we ask for the time measurement to be signal-like. We require that the direction of the lower hemisphere DSA track, as determined by a least-squares fit to the timing in each DT layer
where the fit is not constrained to the IP,
is consistent with being in the downward direction.
We define $t_{\text{DT}}$ as the time at the point of closest approach to the IP as measured by the DTs,
assuming the muon is outgoing. Since cosmic ray muons are incoming in the upper hemisphere
and outgoing in the lower hemisphere, the $t_{\text{DT}}$ of the upper hemisphere track is expected to be 40 to 50\unit{ns} earlier than that of the lower hemisphere
track. As for the signal, since both muons are outgoing, they are reconstructed to have similar times as measured at the IP. Thus, we require that $\Delta t_{\text{DT}}$, which is defined as $\Delta t_{\text{DT}}=t_{\text{DT}}(\text{upper}) -  t_{\text{DT}}(\text{lower})$, is
greater than $-20$\unit{ns}, which greatly reduces the cosmic ray muon background.

In addition to these DT timing variables,
we use a timing measurement from the RPCs that assigns a BX to each hit.
For each of the six layers of the RPCs, the hit is given a BX assignment.
A typical prompt muon created at the IP has a BX assignment of 0 for each of its RPC hits.
The BX assignments of cosmic ray
muons are especially useful in the lower hemisphere of the detector,
as the incoming cosmic ray muons will typically trigger the event and thus be assigned BX values of 0 in
each RPC layer, but the outgoing cosmic ray muons are often assigned positive BX values.
For example, a lower hemisphere cosmic ray muon typically has a BX assignment of 2 for each of its good RPC hits. For the signal, each RPC BX assignment for each muon is typically 0.

Given the BX assignments in each RPC layer for a muon, we can compute the average RPC hit BX assignment
multiplied by 25\unit{ns} as the RPC time for a track ($t_{\text{RPC}}$) and use this as a discriminating variable.
A typical muon from the benchmark decays
has a $t_{\text{RPC}}$ of 0\unit{ns} for both upper and lower hemisphere DSA muon tracks.
On the other hand, the $t_{\text{RPC}}$ of a cosmic ray muon is typically 25 or 50\unit{ns} in the lower hemisphere and
0\unit{ns} in the upper hemisphere. We define $\Delta t_{\text{RPC}}=t_{\text{RPC}}(\text{upper}) - t_{\text{RPC}}(\text{lower})$,
and we require $\Delta t_{\text{RPC}}> -7.5$\unit{ns} to further select signal-like events.

\begin{figure}[hbtp]
\centering
\includegraphics[scale=0.39]{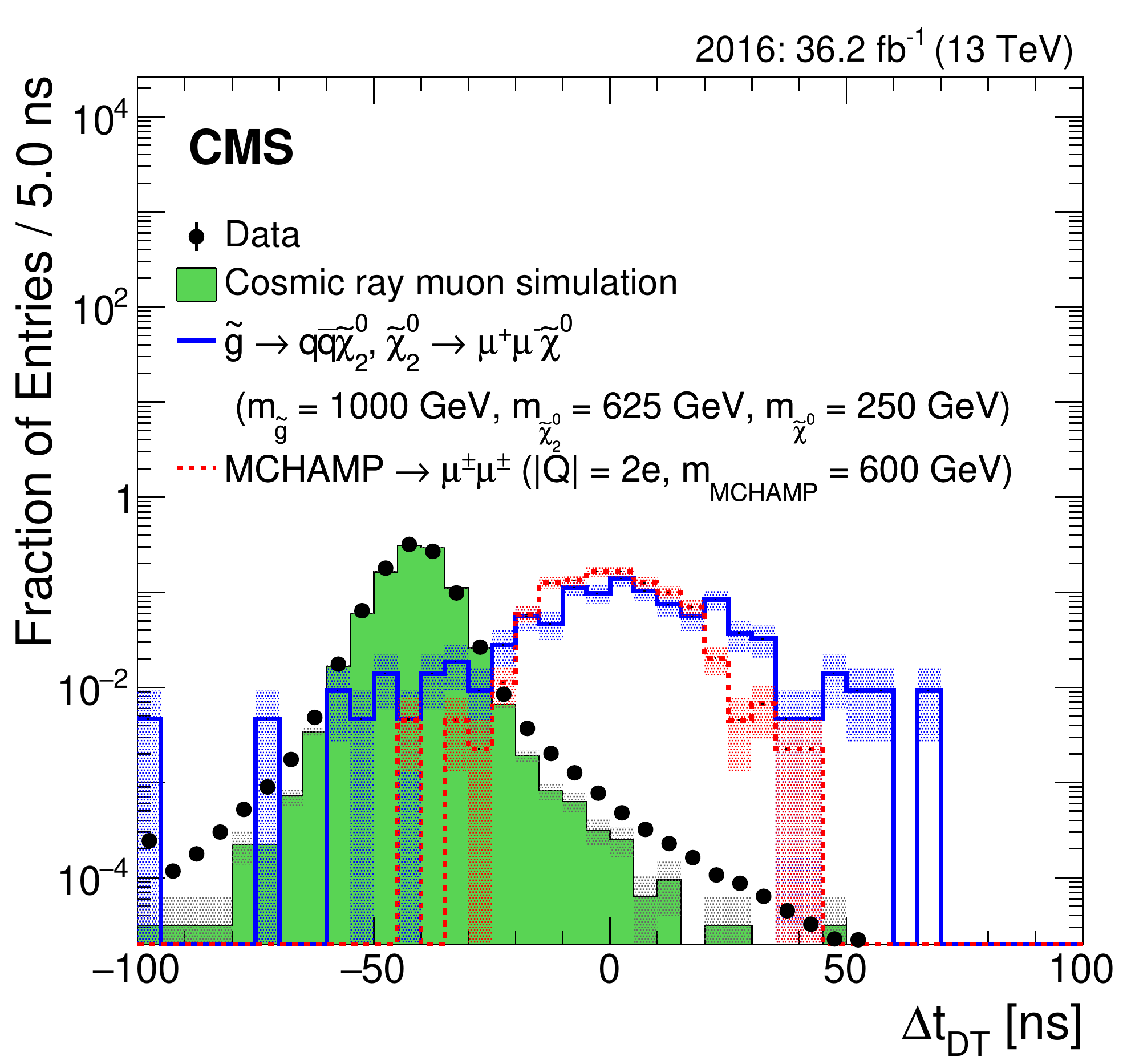}
\includegraphics[scale=0.39]{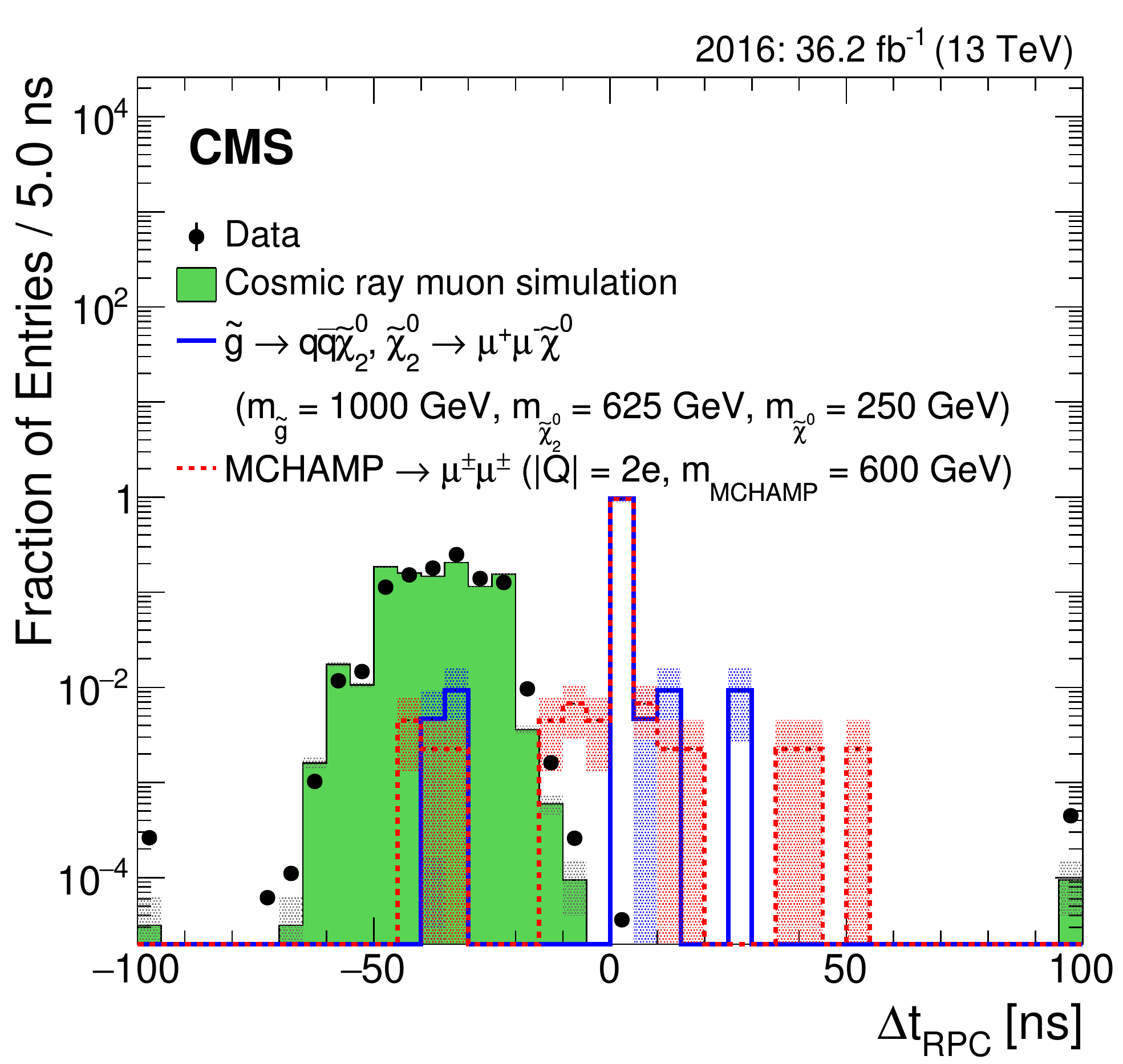}
\centering{}\caption{\label{fig:DSAdeltaDtRpcTiming}
The $\Delta t_{\text{DT}}$ (\cmsLeft) and $\Delta t_{\text{RPC}}$ (\cmsRight) distributions
for 2016 data, MC simulated cosmic ray muon, 1000\GeV gluino signal, and 600\GeV MCHAMP signal events, for the muon search.
The events plotted pass a subset of the full analysis selection that is designed to select good-quality DSA muon tracks but does not reject the cosmic ray muon background.
The number of cosmic ray muon background events is greatly reduced when the full selection is applied, as we require $\Delta t_{\text{DT}}>-20$\unit{ns} and $\Delta t_{\text{RPC}}>-7.5$\unit{ns}.
The gray bands indicate the statistical uncertainty in the simulation.
The histograms are normalized to unit area.
}
\end{figure}

Figure \ref{fig:DSAdeltaDtRpcTiming} shows $\Delta t_{\text{DT}}$ (\cmsLeft) and $\Delta t_{\text{RPC}}$ (\cmsRight) for data and MC simulation.
The events shown here contain good-quality DSA muon tracks, but they are dominated by the cosmic muon background; they are selected with a subset of the criteria described above.
This selection is defined by the
same trigger and reconstructed vertices requirements as above. Additionally, exactly one DSA track in the upper
hemisphere and exactly one DSA track in the lower hemisphere are required. Looser requirements than in the full selection are placed on the DSA track \pt
($>$10\GeV), the number of DT chambers with valid hits (greater than one), and the number of valid RPC
hits (greater than one). We require the same number of DT hits with good timing measurements per DSA track and
number of valid CSC hits as above for this selection. None of the remaining criteria from the main selection criteria described
above are used to select the events in Fig.~\ref{fig:DSAdeltaDtRpcTiming}. As can be seen in Fig.~\ref{fig:DSAdeltaDtRpcTiming}, the number of cosmic ray muon background events is greatly reduced when the full selection is applied, as we require $\Delta t_{\text{DT}}>-20$\unit{ns} and $\Delta t_{\text{RPC}}>-7.5$\unit{ns}.
Since $\Delta t_{\text{DT}}$ and $\Delta t_{\text{RPC}}$ correspond to independent measurements of essentially the same quantity,
a mismeasured cosmic ray muon is much less likely to pass both selections than just one; adding the second requirement improves the rejection of simulated cosmic ray muons by a factor of approximately 350.

\section{Signal efficiency}
\label{sec:signalEfficiency}
In this section, we describe the calculation of the signal efficiency $\varepsilon_{\text{signal}}$, which is the product of several efficiencies.
In the calorimeter search, the stopping efficiency $\varepsilon_{\text{stopping}}$ is the probability that the R-hadron stops in the HB or ECAL barrel (EB), while in
the muon search, $\varepsilon_{\text{stopping}}$ is the probability of each LLP to stop in any region of the detector.
The Stage 1 simulation determines $\varepsilon_{\text{stopping}}$.
The reconstruction efficiency $\varepsilon_{\text{reco}}$ is the efficiency of an event to pass all of the selection
criteria, including the trigger, and it is computed independently of $\varepsilon_{\text{stopping}}$.
In addition, $\varepsilon_{\text{reco}}$ is calculated assuming that the LLP decay occurs when the trigger is live in between collisions,
and assuming a branching fraction ($\mathcal{B}$) of $100\%$ to the decays in the signal models described above.
The Stage 2 simulation determines $\varepsilon_{\text{reco}}$.
The efficiency $\varepsilon_{\text{signal}}$ is defined as the product of $\varepsilon_{\text{stopping}}$ and $\varepsilon_{\text{reco}}$ for the muon search.
For the calorimeter search, $\varepsilon_{\text{signal}}$ is the product of $\varepsilon_{\text{stopping}}$, $\varepsilon_{\text{reco}}$, and two additional factors, $\varepsilon_{\text{CSCveto}}$ and $\varepsilon_{\text{DTveto}}$, which are defined in the next subsection.

\subsection{Calorimeter search} \label{sec:signalEfficiency_caloSearch}

For the calorimeter search, $\varepsilon_{\text{stopping}}$ is constant at about 0.054 for gluinos and 0.045 for top squarks, for the range of masses considered. The $\varepsilon_{\text{stopping}}$ value is larger for gluinos than for top squarks of the same mass because gluinos are more likely to produce doubly charged R-hadrons.

\begin{figure}[!ht]
  \centering
    \includegraphics[width=0.49\textwidth]{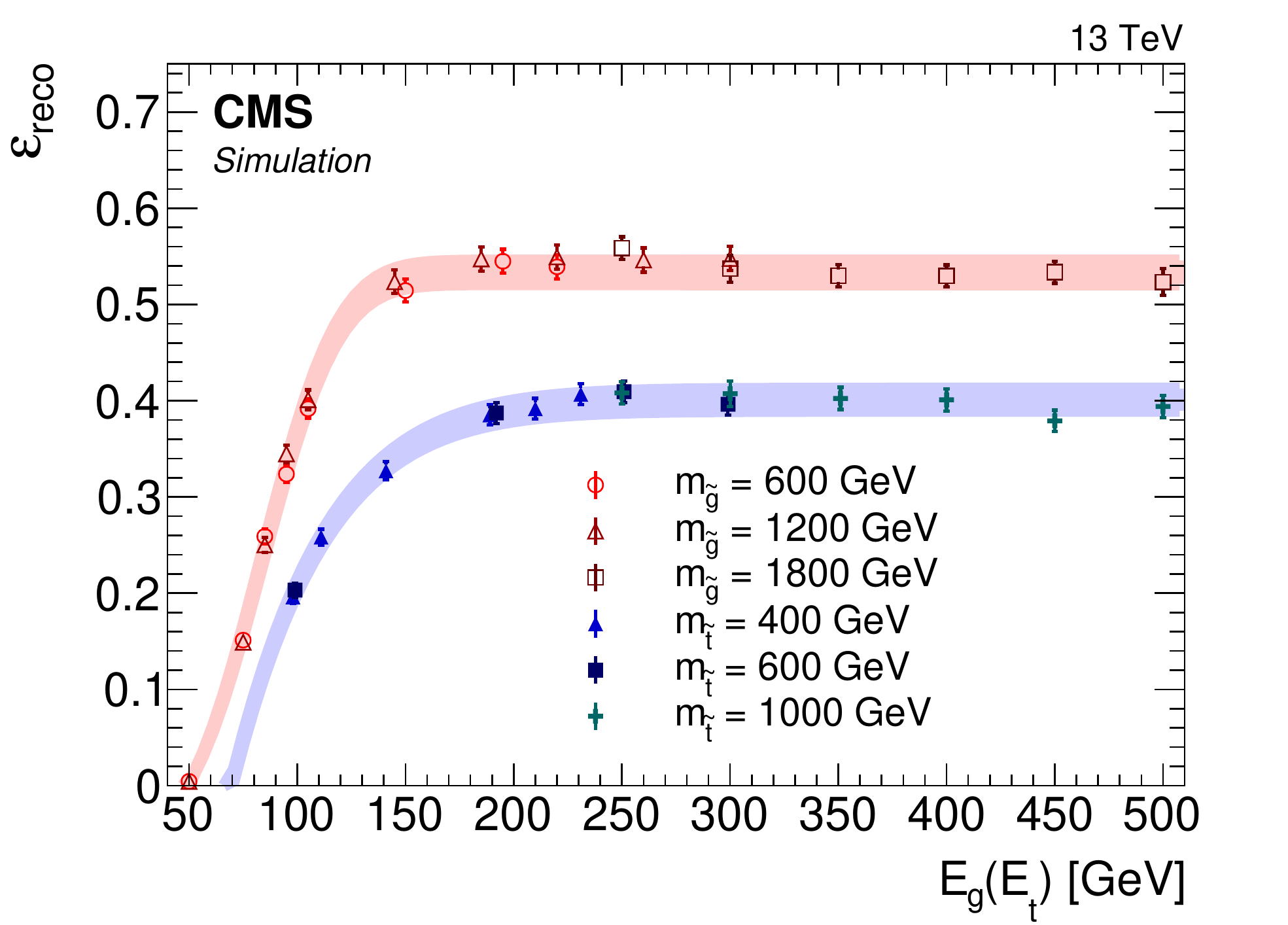}
    \includegraphics[width=0.49\textwidth]{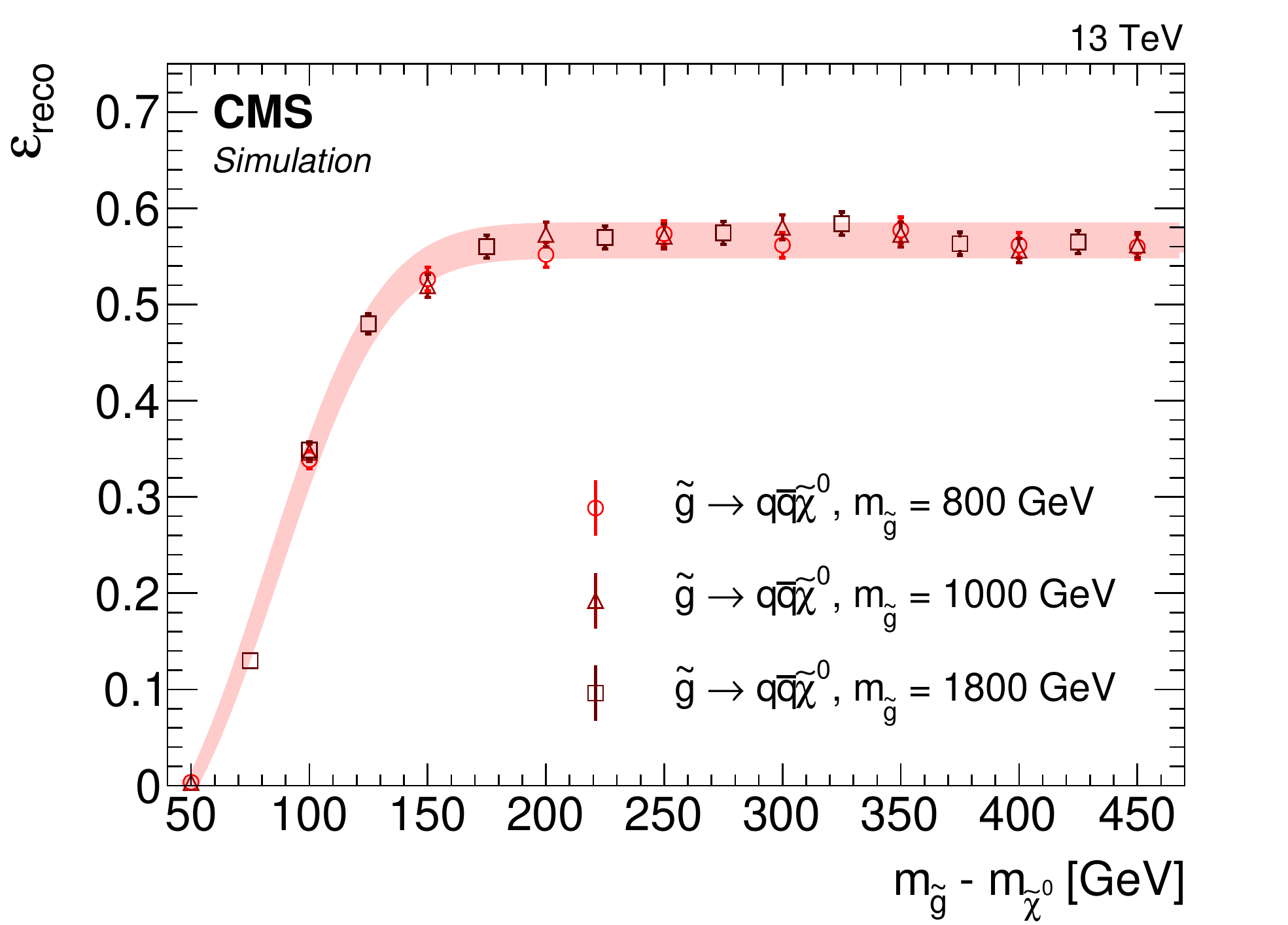}
    \caption{The $\varepsilon_{\text{reco}}$ values as a function of $E_{\cPg}$ or $E_{\cPqt}$ (\cmsLeft), and $m_{\PSg} - m_{\PSGcz}$ (\cmsRight), for $\PSg$ and $\PSQt$ R-hadrons that stop in the EB or HB, in the MC simulation, for the calorimeter search. The $\varepsilon_{\text{reco}}$ values are plotted for the two-body gluino and top squark decays (\cmsLeft) and for the three-body gluino decay (\cmsRight). The shaded bands correspond to the systematic uncertainties, which are described in Section~\ref{sec:systematics}.}
    \label{fig:RecoEff}
\end{figure}

The value of $\varepsilon_{\text{reco}}$ depends primarily on the energy of the visible daughter particle(s) of the R-hadron decay, denoted by $E_{\cPg}$ ($E_{\cPqt}$) if the daughter is a gluon (top quark). When $E_{\cPg} > 130 \GeV$ ($E_{\cPqt} > 170\GeV$), $\varepsilon_{\text{reco}}$ becomes approximately constant, as shown in Fig.~\ref{fig:RecoEff}. For the three-body gluino decay, $\varepsilon_{\text{reco}}$ depends approximately on the mass difference between $\PSg$ and $\PSGcz$, becoming constant when $m_{\PSg} - m_{\PSGcz} \gtrsim 160 \GeV$.

Some physical effects that are not modeled in simulation can cause reconstructed CSC or DT segments that are out of time with respect to a collision. For example, thermal neutrons can take up to a tenth of a second after being produced in $\Pp\Pp$ collisions before they arrive at the muon detectors and induce a signal in the CSCs or DTs. Since these segments can occur when the trigger is live, it is possible that some of the events in the search sample could contain such segments. These events would be rejected by the selection criteria, thus decreasing the probability for a signal to be observed. The terms $\varepsilon_{\text{CSCveto}}$ and $\varepsilon_{\text{DTveto}}$ measure this decrease in efficiency due to these sources.

\begin{table}
  \centering
  \topcaption{Summary of the values of $\varepsilon_{\text{stopping}}$, $\varepsilon_{\text{CSCveto}}$, $\varepsilon_{\text{DTveto}}$, and the plateau value of $\varepsilon_{\text{reco}}$ for different signals, for the calorimeter search. The efficiency $\varepsilon_{\text{stopping}}$ is constant for the range of signal masses considered. The efficiency $\varepsilon_{\text{reco}}$ is given on the $E_{\cPg}$ or $E_{\cPqt}$ plateau for each signal.}
  \begin{tabular}{lccc}
  \hline \\[-2.4ex]
     &  $\PSg \to \cPg\PSGcz$ &  $\PSg \to \cPq\cPaq\PSGcz$ &  $\PSQt \to \cPqt\PSGcz$\\
    \hline
    $\varepsilon_{\text{stopping}}$    & 0.054  & 0.054  & 0.045\\
    $\varepsilon_{\text{reco}}$        & 0.533  & 0.566  & 0.399\\
    $\varepsilon_{\text{CSCveto}}$     & 0.944  & 0.944  & 0.944\\
    $\varepsilon_{\text{DTveto}}$      & 0.877  & 0.877  & 0.877\\
    \hline
    $\varepsilon_{\text{signal}}$      & 0.023  & 0.025  & 0.014\\
    \hline
\end{tabular}
\label{tbl:summary_efficiency}
\end{table}

We define $\varepsilon_{\text{CSCveto}}$ ($\varepsilon_{\text{DTveto}}$) as the conditional probability that a signal passes the beam halo (cosmic ray muon) rejection criteria assuming the potential occurrence of coincident CSC (DT) segments, given that the signal itself passes the full selection criteria. HCAL noise events that are collected by the trigger are used to estimate these two efficiencies from data, since this noise is independent of any muon detector activities and should pass both beam halo rejection and cosmic ray muon rejection criteria. These events are selected by inverting some of the noise rejection criteria. Then $\varepsilon_{\text{CSCveto}}$ ($\varepsilon_{\text{DTveto}}$) is simply the percentage of noise events that survive the beam halo (cosmic ray muon) vetoes among all selected noise events.

Table~\ref{tbl:summary_efficiency} summarizes the values of $\varepsilon_{\text{stopping}}$, $\varepsilon_{\text{CSCveto}}$, $\varepsilon_{\text{DTveto}}$, and the plateau value of $\varepsilon_{\text{reco}}$.

\subsection{Muon search}
Tables \ref{tab:gluino2016EventYieldResults} and \ref{tab:mchamp2016EventYieldResults} show $\varepsilon_{\text{stopping}}$ and $\varepsilon_{\text{reco}}$ for each assumed signal mass in the muon search. The $\varepsilon_{\text{signal}}$ value is the product of these two efficiencies. The $\varepsilon_{\text{stopping}}$ value is larger for MCHAMPs than for gluinos because the MCHAMPs considered have $\abs{Q}=2e$ and the gluinos sometimes produce singly charged R-hadrons. We lose signal efficiency
because the L1 muon trigger is designed to identify muons coming from the IP,
although the muons from the signal can be very displaced.
A further loss in signal efficiency is due to the very strict requirements on the quality of the DSA muon track.
Similarly, the requirement to have
exactly one DSA track traversing the upper hemisphere and exactly one DSA track traversing the lower hemisphere
further reduces the geometrical acceptance,
particularly for the gluino decay, which does not produce back-to-back muons, unlike the MCHAMP decay.
The numbers in Tables \ref{tab:gluino2016EventYieldResults} and \ref{tab:mchamp2016EventYieldResults} represent the maximum number of signal events that can be measured before applying the different search windows depending on the lifetime
of the stopped particle.

\begin{table}[htb]
\centering
\topcaption{Gluino $\varepsilon_{\text{stopping}}$ and $\varepsilon_{\text{reco}}$, as well as the
number of expected gluino events with lifetimes between 10\mus and 1000\unit{s}, assuming $\mathcal{B}(\PSg \to \cPq\cPaq\PSGczDt) \mathcal{B}(\PSGczDt \to \MM\PSGcz)=100\%$, for each mass point considered for the 2016 muon search. The efficiencies are constant for this range of lifetimes.}
\label{tab:gluino2016EventYieldResults}
\begin{tabular}{cccc}
\hline
$m_{\PSg}$ [\GeVns{}]& $\varepsilon_{\text{stopping}}$ & $\varepsilon_{\text{reco}}$ & Expected events\\
\hline
400  &  0.19 & 0.0015 & 400 \\
600  &  0.17 & 0.0024 & 50 \\
800  &  0.17 & 0.0037 & 10 \\
1000 &  0.17 & 0.0029 & 2 \\
1200 &  0.18 & 0.0025 & 0.5 \\
1400 &  0.20 & 0.0031 & 0.2 \\
1600 &  0.21 & 0.0029 & 0.1 \\
\hline
\end{tabular}
\end{table}

\begin{table}[htb]
\centering
\topcaption{MCHAMP $\varepsilon_{\text{stopping}}$ and $\varepsilon_{\text{reco}}$, as well as the
number of expected MCHAMP events with lifetimes between 10\mus and 1000\unit{s}, assuming $\mathcal{B}(\text{MCHAMP} \to \mu^{\pm}\mu^{\pm})=100\%$,
for each mass point considered for the 2016 muon search. The efficiencies are constant for this range of lifetimes.}
\label{tab:mchamp2016EventYieldResults}
\begin{tabular}{cccc}
\hline
$m_{\text{MCHAMP}}$ [\GeVns{}] & $\varepsilon_{\text{stopping}}$ & $\varepsilon_{\text{reco}}$ & Expected events\\
\hline
100   & 0.33 & 0.0059 & 100\\
200   & 0.29 & 0.041  & 50\\
400   & 0.28 & 0.045  & 4\\
600   & 0.25 & 0.042  & 0.5\\
800   & 0.30 & 0.038  & 0.1\\
\hline
\end{tabular}
\end{table}

\section{Background estimation}
Since the background sources in both the calorimeter and the muon searches are not well modeled in simulation, we use control samples in data to estimate their
contributions after the full event selection is applied.

\subsection{Calorimeter search}
After applying the selection criteria in the calorimeter search, some background sources from cosmic ray muons, beam halo, and calorimeter noise remain in the data. We quantify the probability of background events escaping the background vetoes and thus being observed by this search. These inefficiencies are calculated as follows.

We generate a sample of cosmic ray muon events to estimate the rate of such events escaping the cosmic ray muon rejection criteria. The events are generated using \textsc{cmscgen}~\cite{Biallass:2009ev}, a generator based on the air shower program \textsc{corsika}~\cite{Heck:1998vt} and validated in a CMS analysis~\cite{Hesketh:2010iw}. We require that the events pass the preselection criteria, namely that they are required to have substantial energy deposits in the calorimeter and no CSC segments in the muon endcap system. The cosmic ray muon veto inefficiency is defined as
the fraction of preselected simulated cosmic ray muon events that are not rejected by the cosmic ray muon rejection criteria. It is found to be $1\times 10^{-3}$.
To account for the small difference in occupancy between the cosmic ray muon events in data and MC simulation, we first bin the simulated events in the number of DT and outer barrel RPC hits and calculate the inefficiency bin by bin. Then, we apply the halo veto and the noise veto to a sample of events in data, and bin these data events in the same way as the simulated events. For each bin, we multiply the inefficiency by the number of events in data, giving the binned cosmic ray muon prediction. The nominal cosmic ray muon background prediction is then the sum of the events in each bin.

The uncertainty in the cosmic ray muon background is due to the uncertainty in the estimate of muons that escape detection by passing through uninstrumented regions of the CMS detector, which is necessarily estimated from simulation. Since data in the uninstrumented regions are {\it ipso facto} not available to compare to simulation, we define equivalent fiducial volumes of instrumented regions of the muon system. Using these as a proxy for the uninstrumented regions, we assess the reliability of the simulation by comparing data and simulation. We find the average discrepancy between cosmic ray muon data and simulation in the number of detected muons traveling through various fiducial regions in the detector to be about 32\%, and we assign this to be the systematic uncertainty in the cosmic ray muon background estimate. Thus, we estimate the cosmic ray muon background to be $2.6 \pm 0.9$\,($8.8 \pm 3.1$) events in 2015\,(2016) data.

Because there was a high rate of beam halo production in 2015 and 2016 data, and because it is possible for halo muons to escape the acceptance of the endcap muon system, the halo background is nonnegligible.
We estimate the halo veto inefficiency using a tag-and-probe method~\cite{inclusiveWandZcrosssections} that analyzes a high-purity sample of halo events by selecting events having one calorimeter jet with $\abs{\eta} < 1.0$ and CSC segments in at least two endcap layers of the muon system. Since the rates of beam halo in each beam are not the same, the events are first classified according to whether they originated in the clockwise ($-z$ direction) or the counterclockwise ($+z$ direction) beam. Then for each class, depending on whether these events have CSC segments in only one endcap or both endcaps of the muon system, they are categorized into events that have only the incoming portion of a halo muon track, events that have only the outgoing portion, and events that have both portions. The number of events that escape detection is $N_{\text{IncomingOnly}} N_{\text{OutgoingOnly}}/N_{\text{Both}}$. We define $N_{\text{IncomingOnly}}$ ($N_{\text{OutgoingOnly}}$) as the number of events that have only an incoming (outgoing) portion of a halo muon track. The number of events that have both an incoming and an outgoing halo muon track is $N_{\text{Both}}$. After binning halo events in their $x$ and $y$ coordinates and performing the classification and calculation discussed above, we estimate the halo veto inefficiency to be $1\times 10^{-4}$. We then multiply this inefficiency by the number of halo events vetoed in the search region.

To account for the possibility that the $x$-$y$ binning does not reproduce the actual shape of the inactive or uninstrumented regions of the detector, thus biasing the estimate, we repeat the calculation above, but binning events in $\phi$ and $r$ instead. The systematic uncertainty is then defined as the difference between the results from the two binning schemes. We find a halo background estimate of $1.1 \pm 0.1$\,($2.6 \pm 0.2$) events in 2015\,(2016) data.

Finally, the background estimation of instrumental noise is performed using control data in dedicated cosmic runs with no beams in the LHC, which include only cosmic ray muon and noise events. We select cosmic runs taken several days after $\Pp\Pp$ collision runs so that there would be little chance for the signal to appear. After applying all selection criteria on the control data, we observe 2 events in each of the 2015 and 2016 control data. We then subtract the expected cosmic ray muon background from the total event yield, obtaining a noise background estimate of $0.3^{+2.4}_{-0.3}$ ($0.0^{+2.2}_{-0.0}$) events in 2015 (2016) control data. Based on the number of noise events in the control sample, we expect the noise veto inefficiency to be $\leq 1\times 10^{-4}$. These noise estimates are then scaled to the search data, assuming that the noise veto inefficiency remains the same. The resulting noise background estimate is $0.4^{+2.9}_{-0.4}$ ($0.0^{+9.8}_{-0.0}$) events in 2015\,(2016). The uncertainty in the 2016 prediction is large because the trigger livetime of the cosmic runs in 2016 was about 60\% shorter than that of the collision runs, and also because the 2016 trigger livetime in collision runs is larger than the 2015 trigger livetime. Therefore, the uncertainty is scaled by a larger factor.

The total background estimate for the calorimeter search is $4.1^{+3.0}_{-1.0}$ ($11.4^{+10.3}_{-3.1}$) events in 2015 (2016), as summarized in Table~\ref{tbl:bkg2015search}.

\begin{table}[!ht]
  \centering
  \topcaption{The background prediction for the calorimeter search. The total background median value is listed in parentheses; this value corresponds directly to the median expected limits shown below.}
  \begin{tabular}{lcccccc}
    \hline
    LHC         &Trigger          &HCAL    &Cosmic ray &Beam  &Total\\
    period      &livetime [hrs]   &noise   &muons      &halo  &background \\
    \hline \\[-2.4ex]
    2015        &135    & $0.4^{+2.9}_{-0.4}$ & $2.6 \pm 0.9$ & $1.1 \pm 0.1$  & $4.1^{+3.0}_{-1.0}$ (6.2)\\[0.2ex]
    2016        &586    & $0.0^{+9.8}_{-0.0}$ & $8.8 \pm 3.1$ & $2.6 \pm 0.2$  & $11.4^{+10.3}_{-3.1}$ (17.4) \\[0.2ex]
    \hline
  \end{tabular}
  \label{tbl:bkg2015search}
\end{table}

\subsection{Muon search}
\begin{figure}[hbtp]
\centering
\includegraphics[width=0.6\textwidth]{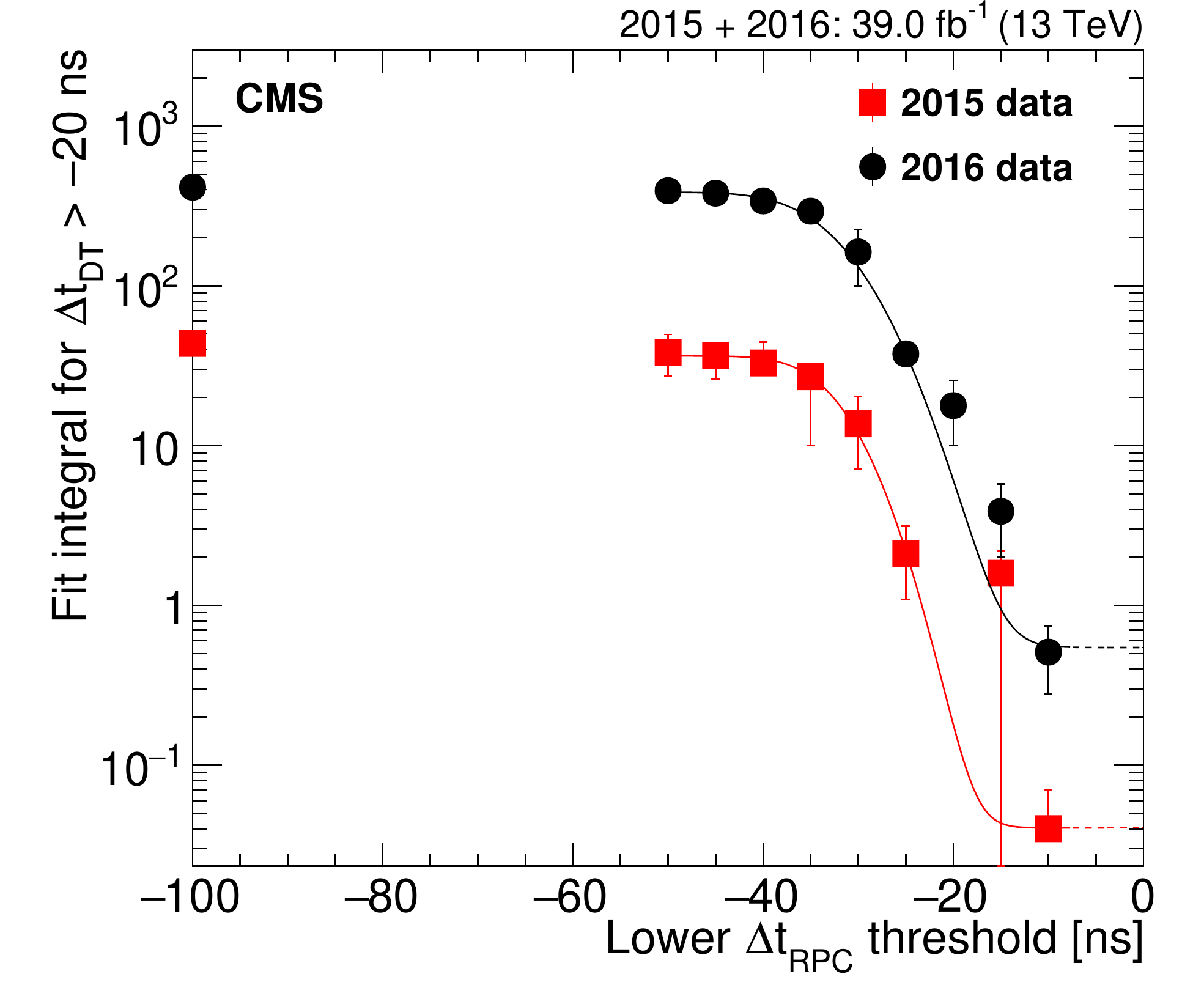}
\caption{\label{fig:backgroundExtrapolation_2GausCrystalBall_errorFunctionFit}
The background extrapolation for the muon search.
The integral of the fit function to $\Delta t_{\text{DT}}$ with the sum of two Gaussian distributions and a Crystal Ball function, for $\Delta t_{\text{DT}}>-20$\unit{ns},
is plotted as a function of the lower $\Delta t_{\text{RPC}}$ selection, for 2015 (red squares) and 2016 (black circles) data.
The points are fitted with an error function and used to extrapolate to the signal region, which is
defined as $\Delta t_{\text{RPC}}>-7.5$\unit{ns}.
}
\end{figure}

In the muon search, a small number of cosmic ray muon background events remains after applying the full event selection to the data. The cosmic ray muon background is estimated by extrapolating the data from a background-dominated region into the signal region.
We apply the full event selection to the data except the $\Delta t_{\text{DT}}$ criterion and invert the $\Delta t_{\text{RPC}}$ criterion.
We then fit the $\Delta t_{\text{DT}}$ distribution with the sum of two Gaussian distributions and a Crystal Ball function~\cite{CBfunction},
since $\Delta t_{\text{DT}}$ is relatively Gaussian with a long asymmetrical tail.
Next, we compute the integral of the fit function, for $\Delta t_{\text{DT}}>-20$\unit{ns}. Then, we compute the same integral after having tightened the selection
criteria on $\Delta t_{\text{RPC}}$ to $-50 < \Delta t_{\text{RPC}} < -7.5$\unit{ns}, then $-45 <\Delta t_{\text{RPC}} < -7.5$\unit{ns},
etc.\,in steps of 5\unit{ns} up to $-10 <\Delta t_{\text{RPC}}< -7.5$\unit{ns}.
Finally, we plot each integral as a function of
the lower selection on $\Delta t_{\text{RPC}}$, and fit this with an error function to extrapolate to the
$\Delta t_{\text{RPC}}>-7.5$\unit{ns} region (see Fig.~\ref{fig:backgroundExtrapolation_2GausCrystalBall_errorFunctionFit}).
We use an error function fit in order to make a conservative background estimate.
Given this extrapolation, we predict
0.04 background events in 2015 data, with a negligible statistical uncertainty, and
$0.50\pm0.02$ background events in 2016 data, where the uncertainty given is statistical only.
The statistical uncertainty in the background prediction derives from the uncertainty in the error function fit parameters.
We checked the background prediction method by repeating the procedure with nonoverlapping $\Delta t_{\text{RPC}}$ regions
and found that the numbers of background events predicted are consistent with the nominal values.

The systematic uncertainty in the background prediction is evaluated by repeating the steps above, except
changing the fit of the $\Delta t_{\text{DT}}$ distribution to the sum of two Gaussian distributions and a Landau function~\cite{Landau:1944if}.
Using the error function fits to extrapolate to $\Delta t_{\text{RPC}}>-7.5$\unit{ns}
gives a prediction of
$0.07\pm0.06$\,($0.10\pm0.01$) background events in 2015\,(2016), where the uncertainty given is statistical only.
Thus, the background prediction is:
$0.04 \pm 0.03$\syst background events in 2015 data, with a negligible statistical uncertainty, and
$0.50 \pm 0.02\stat \pm 0.40\syst$ background events in 2016 data.

Despite the fact that we require exactly one upper hemisphere DSA track and exactly one lower
hemisphere DSA track, there could still be some background from two coincident cosmic ray muons. This
background from two coincident cosmic ray muons could occur if the upper hemisphere DSA track of one
cosmic ray muon is reconstructed and if the lower hemisphere DSA track of the other is also reconstructed.
We estimate this contribution from data by finding the rate of events with exactly one reconstructed
DSA track in one hemisphere satisfying all of the selection criteria except for the $\Delta t_{\text{DT}}$
and $\Delta t_{\text{RPC}}$ criteria, and no tracks in the other hemisphere.
Then, making simple assumptions about when the two coincident cosmic ray muons could occur and about the DSA track
reconstruction efficiency as a function of BX, we calculate the number of accidentally coincident cosmic ray
muons and find it to be negligible.
\section{Systematic uncertainties in the signal efficiency}
\label{sec:systematics}
While the \GEANTfour simulation
used to derive the stopping probability accurately
models both the electromagnetic and nuclear interaction
energy loss mechanisms, the relative contributions of
these energy loss mechanisms to the stopping probability
depend significantly on unknown R-hadron spectroscopy.
We do not consider this dependence to be a source of uncertainty for either the calorimeter or the muon search,
however, since for any given model the
resultant uncertainty in the stopping probability is small.
Nevertheless, there are several sources of uncertainty in the signal efficiency measurement.

\subsection{Calorimeter search}
In the calorimeter search, the systematic uncertainty due to the trigger efficiency is negligible since the offline jet energy criterion ensures the data analyzed are well above the turn-on region, so $\varepsilon_{\text{reco}}$ is constant. We consider possible systematic uncertainties in $\varepsilon_{\text{CSCveto}}$ and $\varepsilon_{\text{DTveto}}$ by varying the criteria used to select HCAL noise events that were described in Section~\ref{sec:signalEfficiency_caloSearch}. We compare the efficiency of data events to pass these new HCAL noise criteria with that of the nominal HCAL noise selection criteria, and we find that the relative change in the efficiencies is less than 0.2\% for both $\varepsilon_{\text{CSCveto}}$ and $\varepsilon_{\text{DTveto}}$, and therefore negligible. The uncertainty in the integrated luminosity is estimated as 2.3\,(2.5)\% for 2015\,(2016) data~\cite{CMS:2016eto,CMS:2017sdi}. The relative uncertainty in $\varepsilon_{\text{reco}}$ is estimated to be 7.7\,(5.2)\% for $\PSg$\,($\PSQt$) in the 2015 analysis, and 7.5\,(5.2)\% for $\PSg$\,($\PSQt$) in the 2016 analysis. This uncertainty, which is shown by the shaded bands in Fig.~\ref{fig:RecoEff}, is determined by computing the maximal relative difference among points on the plateau.

Jets in this analysis are not formed by particles originating from the center of the detector, so the standard uncertainty in the jet energy scale does not apply.
Instead, we refer to a study performed on the HCAL during cosmic data taking in 2008~\cite{HCALperformance_cosmicsAndBeamData}. This study compares the energy of the reconstructed jets in simulated cosmic ray muon events and cosmic ray muon events in data, concluding that the uncertainty in the jet energy in the simulation is about 2\%.
Moreover, a study conducted with 2012 data~\cite{1742-6596-587-1-012004} compares the data and simulation for dijets originating from the interaction point. The comparison leads to an estimate of $<$2\% for jets striking the HCAL barrel with angles of incidence from 0 to $\pi/3$. After rescaling the jet energy by 2\%, the signal efficiency varies by 2\%. This estimate is conservative since only the yield of signals with jet energy near the offline threshold is affected by the variation of the jet energy, and as a result the uncertainty decreases rapidly as $E_{\cPg}$ ($E_{\cPqt}$) increases.

We have also considered the uncertainty associated with the jet energy resolution.  Studies have shown that the signal yield is insensitive to variations in this uncertainty, and thus that the systematic uncertainty associated with the jet energy resolution is negligible.

The total systematic uncertainty in the signal yield is 8.3\,(8.2)\% in the 2015\,(2016) search. The systematic uncertainties are summarized in Table~\ref{tbl:syst_uncert}.

\begin{table}[!ht]
  \centering
  \topcaption{Systematic uncertainties in the signal efficiency in the 2015 and 2016 calorimeter searches.}
  \begin{tabular}{lcc}
    \hline
    Systematic uncertainty &2015 &2016 \\
    \hline
    Reconstruction efficiency   &7.7\% &7.5\%\\
    Integrated luminosity       &2.3\% &2.5\%\\
    Jet energy scale            &2.0\% &2.0\%\\
    \hline
  \end{tabular}
  \label{tbl:syst_uncert}
\end{table}

\subsection{Muon search}
The muon search also has several sources of systematic uncertainties.
We consider the systematic uncertainty associated with the MC
simulation modeling of the charge divided by the \pt ($Q/\pt$) resolution by comparing this resolution
in cosmic ray muon data and cosmic ray muon MC simulation. The
resolution compares $Q/\pt$ of the upper and lower hemisphere tracks:
\begin{equation*}
R(Q/\pt)=\frac{(Q/\pt)^{\text{upper}} - (Q/\pt)^{\text{lower}}} {\sqrt{2}(Q/\pt)^{\text{lower}}}.
\end{equation*}

We plot the standard deviation of Gaussian fits of the resolution, as a function of the lower
hemisphere track \pt, for both cosmic ray muon data and MC simulation. A fit of the ratio between data and MC simulation in this plot
for muon tracks in the lower hemisphere with $\pt>50\GeV$ gives a difference between cosmic ray muon data and simulation of
9.0\,(5.3)\% in the 2015\,(2016) analysis. We propagate this resolution uncertainty to an uncertainty in the signal efficiency
by smearing the momentum distribution of muons in the signal and observing the corresponding variation in the signal yields. We take
the largest variation in the signal yield, namely, 13\,(7.0)\% in the 2015\,(2016) analysis, as the systematic uncertainty in the modeling of
the $Q/\pt$ resolution.

There is also a systematic uncertainty associated with the trigger
acceptance. Since the largest difference between data and MC simulation in the plateau of the trigger turn-on curves is
13\,(2.8)\% in the 2015\,(2016) analysis, we take these values as the
systematic uncertainty in the trigger acceptance.

The total systematic uncertainty in the signal yield is 19\,(7.9)\% in the 2015\,(2016) search. The systematic uncertainties are summarized in Table~\ref{tab:systematics}.

\begin{table}[htbp]
\centering
\topcaption{\label{tab:systematics}Systematic uncertainties in the signal efficiency for the 2015 and 2016 muon searches.}
\begin{tabular}{lcc}
\hline
Systematic uncertainty   & 2015   & 2016\\
\hline
$Q/\pt$ resolution mismodeling  & 13\% & 7.0\% \\
Trigger acceptance  & 13\% & 2.8\% \\
Integrated luminosity &  2.3\% & 2.5\% \\
\hline
\end{tabular}
\end{table}
\section{Results}\label{sec:results}
In the calorimeter search, we predict $4.1^{+3.0}_{-1.0}$ ($11.4^{+10.3}_{-3.1}$) background events in the 2015 (2016) data.
Four events that pass all of the selection criteria are observed in 2015 data, while 13 events are observed in 2016 data. Both observed numbers of events are consistent with the predicted backgrounds. The observed events are most likely cosmic ray muon or beam halo events, as they each consist of a single reconstructed jet.

In the muon search, we predict
$0.04\pm0.03$\,($0.50\pm0.40$) background events in 2015 (2016).
There are zero observed events in both 2015 and 2016 data that pass all of the selection criteria.

In both the calorimeter and muon searches, we count the number of observed events in equally spaced $\log_{10}\text{(time)}$ bins of signal lifetime hypotheses from $10^{-7}$ to $10^6$\unit{s}. For lifetime hypotheses shorter than one LHC orbit of $89\mus$, we search within a sensitivity-optimized time window of 1.3 times the stopped particle's lifetime, where the window starts after each $\Pp\Pp$ collision, to avoid the addition of backgrounds for time intervals during which a signal with a given lifetime has a large probability to have already decayed. We assume that the cosmic ray muon background (and noise background in the calorimeter search) is uniformly distributed in time. In the calorimeter search, we estimate the halo background for each lifetime hypothesis by finding the ratio of halo events in the search time window to the total number of halo events, then multiplying this ratio by the halo background estimate for the full trigger livetime. We select the halo events by requiring events to pass all of the selection criteria except the CSC segment veto described above, and then requiring the events to have at least one CSC segment. Then, we determine if these halo events are within the search window by observing how long after the most recent filled BX they occurred.

For lifetimes longer than one orbit, the trigger livetime, the expected background, and the number of observed events are independent of the lifetime. The effective integrated luminosity decreases with lifetime for lifetimes longer than one LHC orbit, and the analysis sensitivity degrades with lifetimes longer than one LHC fill because any signal that decays between fills will have few chances to be observed.

For lifetime hypotheses shorter than one orbit, both the number of observed events and the expected background depend on the time window considered, which is a fraction of the total trigger livetime. Similarly, the effective integrated luminosity is reduced for short lifetimes. As we gradually increase the lifetime in the hypothesis from the minimal value, we include more observed events in the search window. When the lifetime is shorter than one orbit, to explicitly show the discontinuous changes of the upper limits whenever the expanding search window covers a new observed event, we test two lifetime hypotheses in addition to the equally spaced $\log_{10}\text{(time)}$ ones, for each observed event in these counting experiments. These two additional lifetime hypotheses are the largest lifetime hypothesis for which the event lies outside the time window, and the smallest lifetime hypothesis for which the event is contained within the time window.

Tables~\ref{tbl:2016countingExperiment} and~\ref{tab:lifetimeHypotheses_background2016} show the results of the counting experiment for the 2016 data.
The data show no excess over background, and we set upper limits
on the signal production cross section ($\sigma$) using a hybrid method with the $\mathrm{CL_{s}}$
criterion~\cite{Junk_CLS,Read_CLS} to incorporate the systematic uncertainties~\cite{Cousins_systUncertInUpperLimit},
in both the calorimeter and muon searches.
By combining the likelihoods of the search results from the 2015 and 2016 analyses,
we set combined upper limits on $\mathcal{B}\sigma$ for the benchmark signal models.

\begin{table}[!ht]
  \centering
    \topcaption{Counting experiment results for different lifetimes in the calorimeter search with 2016 data.}
  \begin{tabular}{ccccc}
    \hline
    Lifetime [s]    & Effective integrated              & Trigger        & Expected   &Observed\\
                    & luminosity [$\text{fb}^{-1}$]   & livetime [hrs] & background &events\\
    \hline \\[-2.4ex]
    $5\ten{-8}$  & 0.27     &17   &$0.4^{+0.3}_{-0.1}$    &0\\[0.2ex]
    $8\ten{-8}$  & 0.65     &34   &$0.8^{+0.6}_{-0.2}$    &0\\[0.2ex]
    $10^{-7}$         & 1.27     &67   &$1.4^{+1.2}_{-0.4}$    &0\\[0.2ex]
    $10^{-6}$         & 9.98     &417  &$8.4^{+7.5}_{-2.3}$    &8\\[0.2ex]
    $10^{-5}$         & 13.37    &583  &$11.3^{+10.2}_{-3.1}$  &13\\[0.2ex]
    $10^{-4}$         & 13.70    &583  &$11.4^{+10.3}_{-3.1}$  &13\\[0.2ex]
    $10^3$            & 13.57    &583  &$11.4^{+10.3}_{-3.1}$  &13\\[0.2ex]
    $10^4$            & 11.78    &583  &$11.4^{+10.3}_{-3.1}$  &13\\[0.2ex]
    $10^5$            & 8.27     &583  &$11.4^{+10.3}_{-3.1}$  &13\\[0.2ex]
    $10^6$            & 5.61     &583  &$11.4^{+10.3}_{-3.1}$  &13\\[0.2ex]
    \hline
    \end{tabular}
    \label{tbl:2016countingExperiment}
\end{table}

\begin{table}
\centering
\topcaption{\label{tab:lifetimeHypotheses_background2016}
Counting experiment results for different lifetimes in the muon search with 2016 data.}
\begin{tabular}{ccccc}
\hline
    Lifetime [s]    & Effective integrated              & Trigger        & Expected   &Observed\\
                    & luminosity [$\text{fb}^{-1}$]   & livetime [hrs] & background &events\\
\hline \\[-2.4ex]
$5\ten{-8}$  &  0.27 & 11  & $0.01\pm0.01$ & 0  \\
$8\ten{-8}$  &  0.64 & 34  & $0.03\pm0.02$ & 0  \\
$10^{-7}$         &  1.27 & 68  & $0.06\pm0.05$ & 0  \\
$10^{-6}$         &  9.95 & 422 & $0.36\pm0.29$ & 0  \\
$10^{-5}$         & 13.34 & 581 & $0.49\pm0.39$ & 0  \\
$10^{-4}$         & 13.67 & 589 & $0.50\pm0.40$ & 0  \\
1                 & 13.67 & 589 & $0.50\pm0.40$ & 0  \\
$10^{3}$          & 13.55 & 589 & $0.50\pm0.40$ & 0  \\
$10^{4}$          & 11.75 & 589 & $0.50\pm0.40$ & 0  \\
$10^{5}$          &  8.26 & 589 & $0.50\pm0.40$ & 0  \\
$10^{6}$          &  5.61 & 589 & $0.50\pm0.40$ & 0  \\
\hline
\end{tabular}
\end{table}

\begin{figure}[!h]
  \centering
    \includegraphics[width=0.48\textwidth]{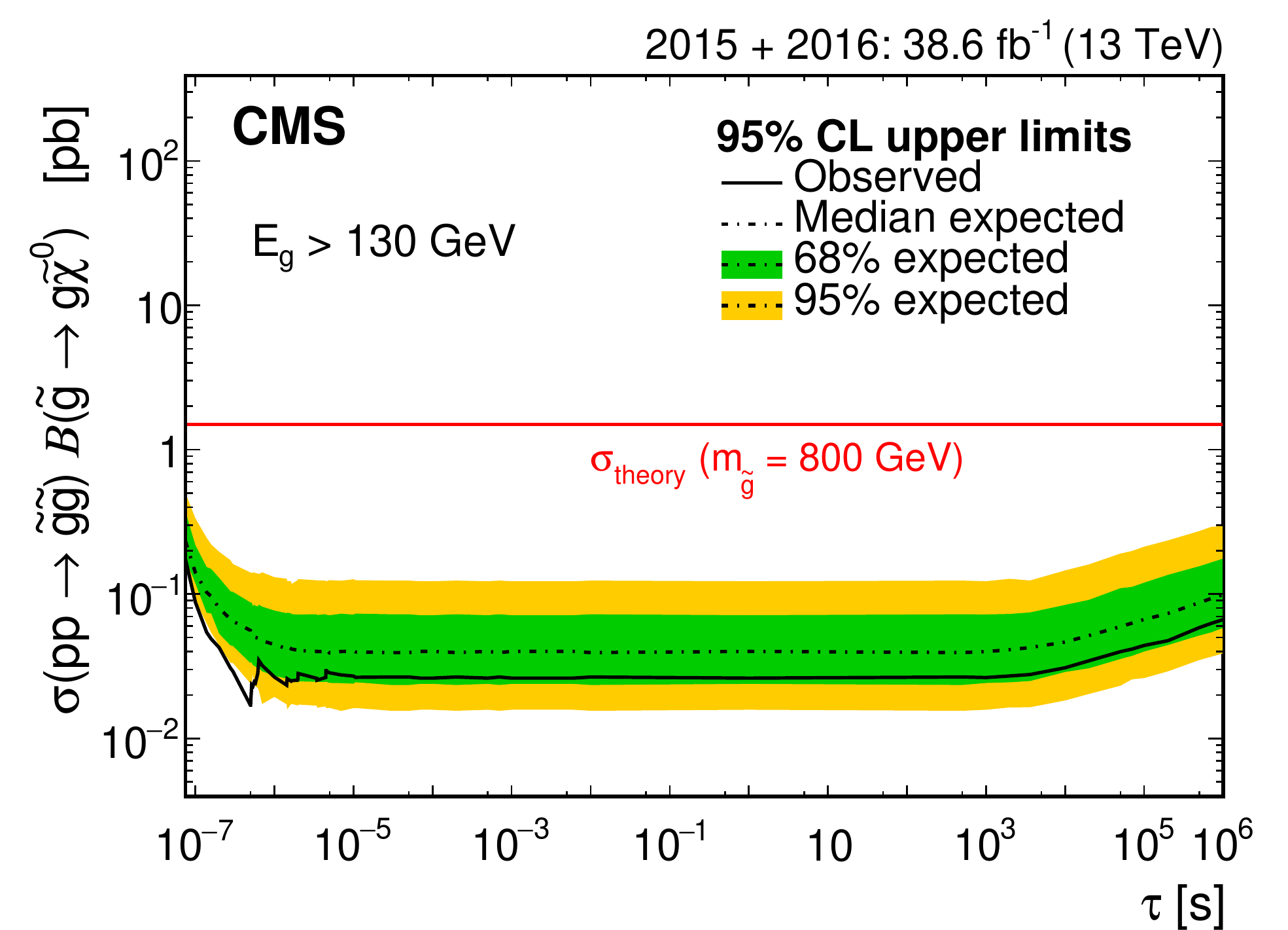}
    \includegraphics[width=0.48\textwidth]{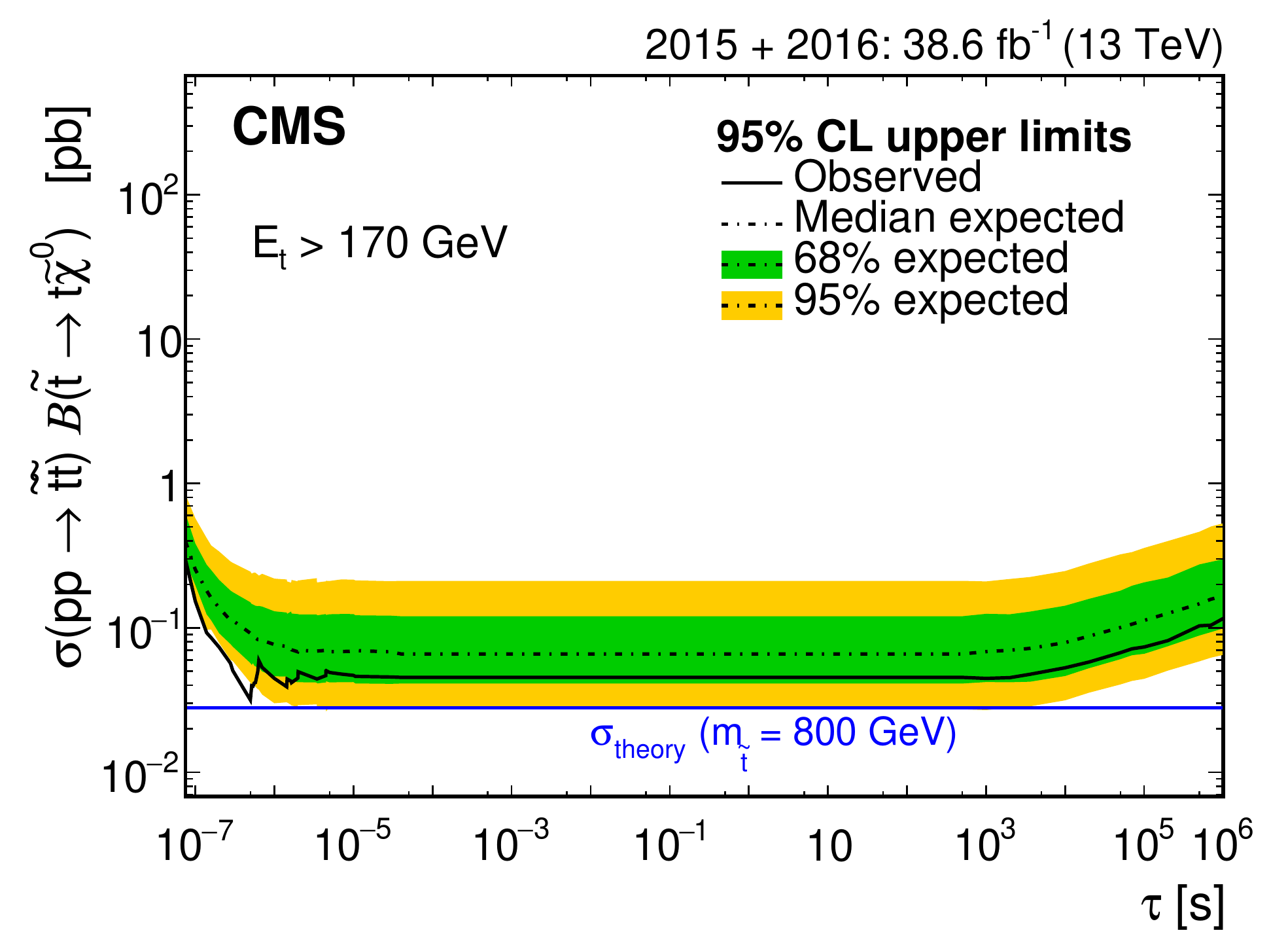}
    \includegraphics[width=0.48\textwidth]{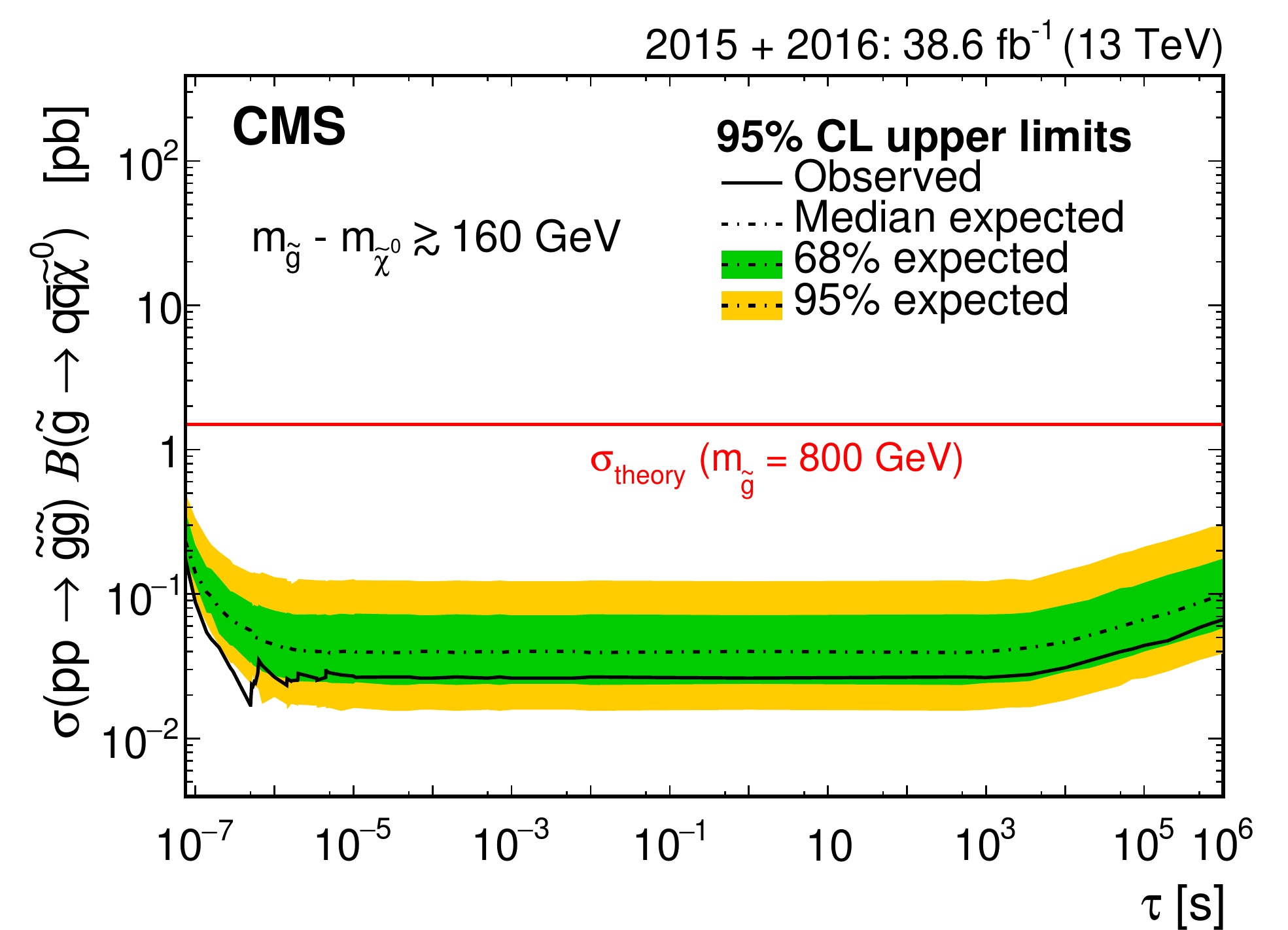}
    \caption{
The 95\% CL upper limits on $\mathcal{B}\sigma$
for gluino and top squark pair production,
using the cloud model of R-hadron interactions,
as a function of lifetime, for combined 2015 and 2016 data for the calorimeter search.
We show gluinos that undergo a two-body decay (upper \cmsLeft), top squarks that undergo a two-body decay (upper \cmsRight), and gluinos that undergo a three-body decay (lower).
The discontinuous structure observed between $10^{-7}$ and $10^{-5}$ s is due to the increase of the number of observed events in the search window as the lifetime increases.
The theory lines assume $\mathcal{B}=100\%$.
}

    \label{fig:CombinedAllInOneLifetime}
\end{figure}

\begin{figure}[!h]
  \centering
    \includegraphics[width=0.48\textwidth]{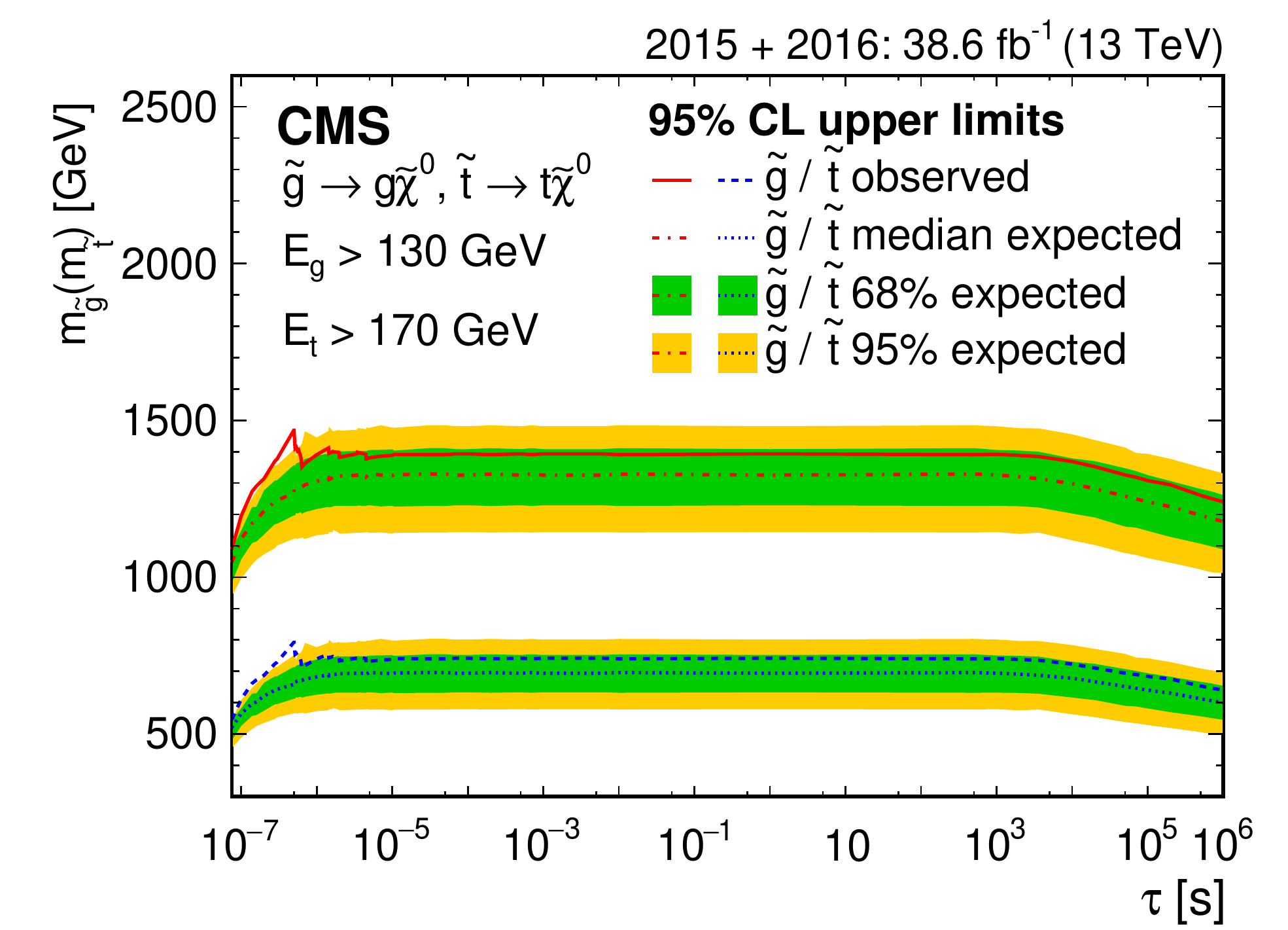}
    \includegraphics[width=0.48\textwidth]{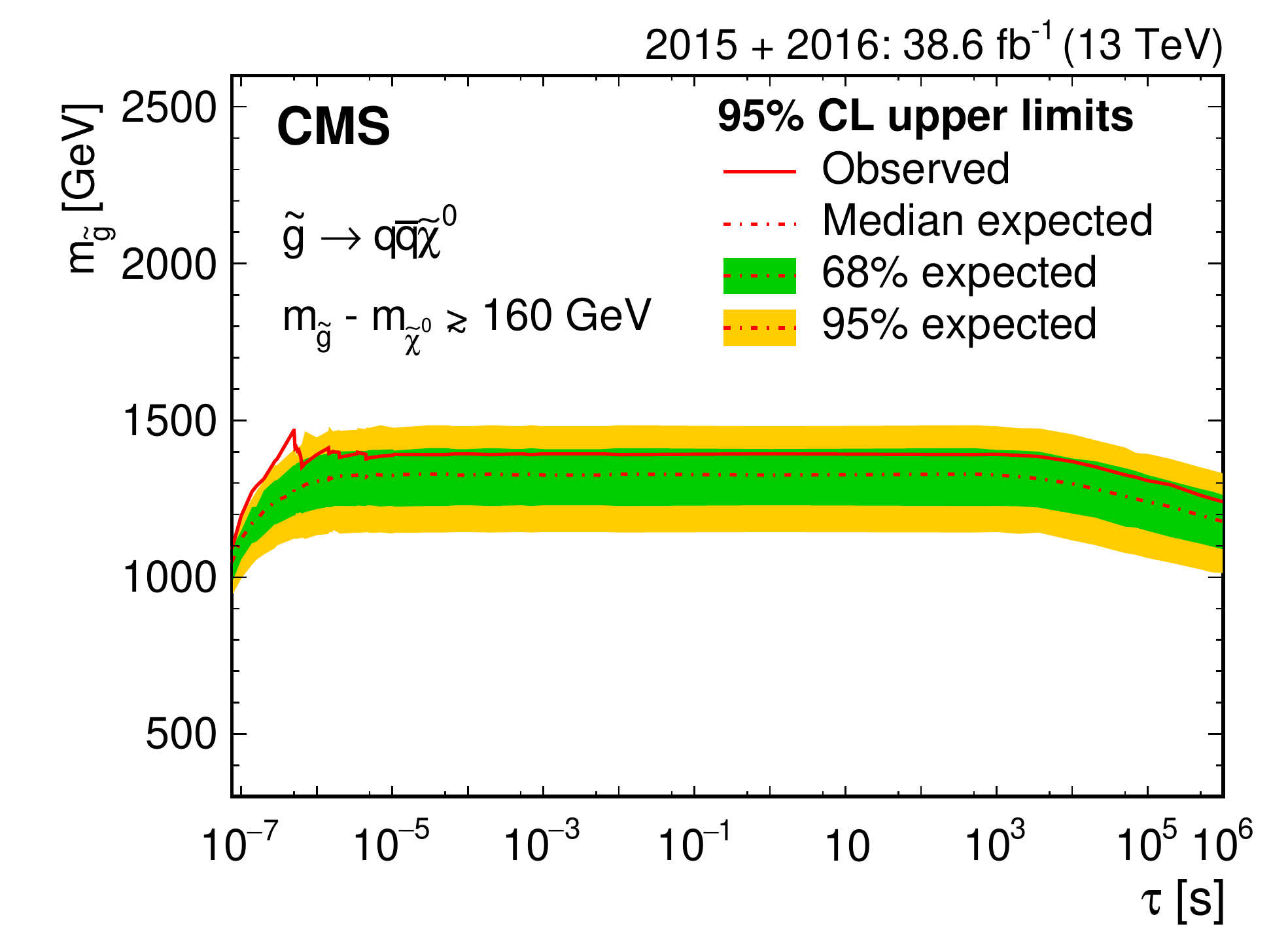}
    \caption{
The 95\% CL upper limits on the gluino and top squark mass,
using the cloud model of R-hadron interactions,
as a function of lifetime, for combined 2015 and 2016 data for the calorimeter search.
We show gluinos and top squarks that undergo a two-body decay (\cmsLeft) and gluinos that undergo a three-body decay (\cmsRight).
The discontinuous structure observed between $10^{-7}$ and $10^{-5}$ s is due to the increase of the number of observed events in the search window as the lifetime increases.
}
    \label{fig:CombinedGluinostopMassLifetime}
\end{figure}

In the calorimeter search, the 95\% confidence level (CL) upper limits on $\mathcal{B}\sigma$ for $\PSg$ ($\PSQt$) pair production for combined 2015 and 2016 data as a function of the particle's lifetime $\tau$ are shown in Fig.~\ref{fig:CombinedAllInOneLifetime}, assuming $E_{\cPg} > 130\GeV$ ($m_{\PSg}-m_{\PSGcz} \gtrapprox 160\GeV$ or $E_{\cPqt} > 170\GeV$). In Fig.~\ref{fig:CombinedGluinostopMassLifetime}, the gluino and top squark mass limits are shown, assuming $\mathcal{B}(\PSg \to \cPg\PSGcz) =\mathcal{B}(\PSg \to \cPq\cPaq\PSGcz)=\mathcal{B}(\PSQt \to \cPqt\PSGcz) = 100\%$. We exclude gluinos with $m_{\PSg} < 1385$\,(1393)\GeV that decay via $\PSg \to \cPg\PSGcz$\,($\PSg \to \cPq\cPaq\PSGcz$) and top squarks with $m_{\PSQt} < 744\GeV$ at 95\% CL for $10\mus < \tau < 1000$\unit{s}.

\begin{figure}[!htb]
  \centering
    \includegraphics[width=0.48\textwidth]{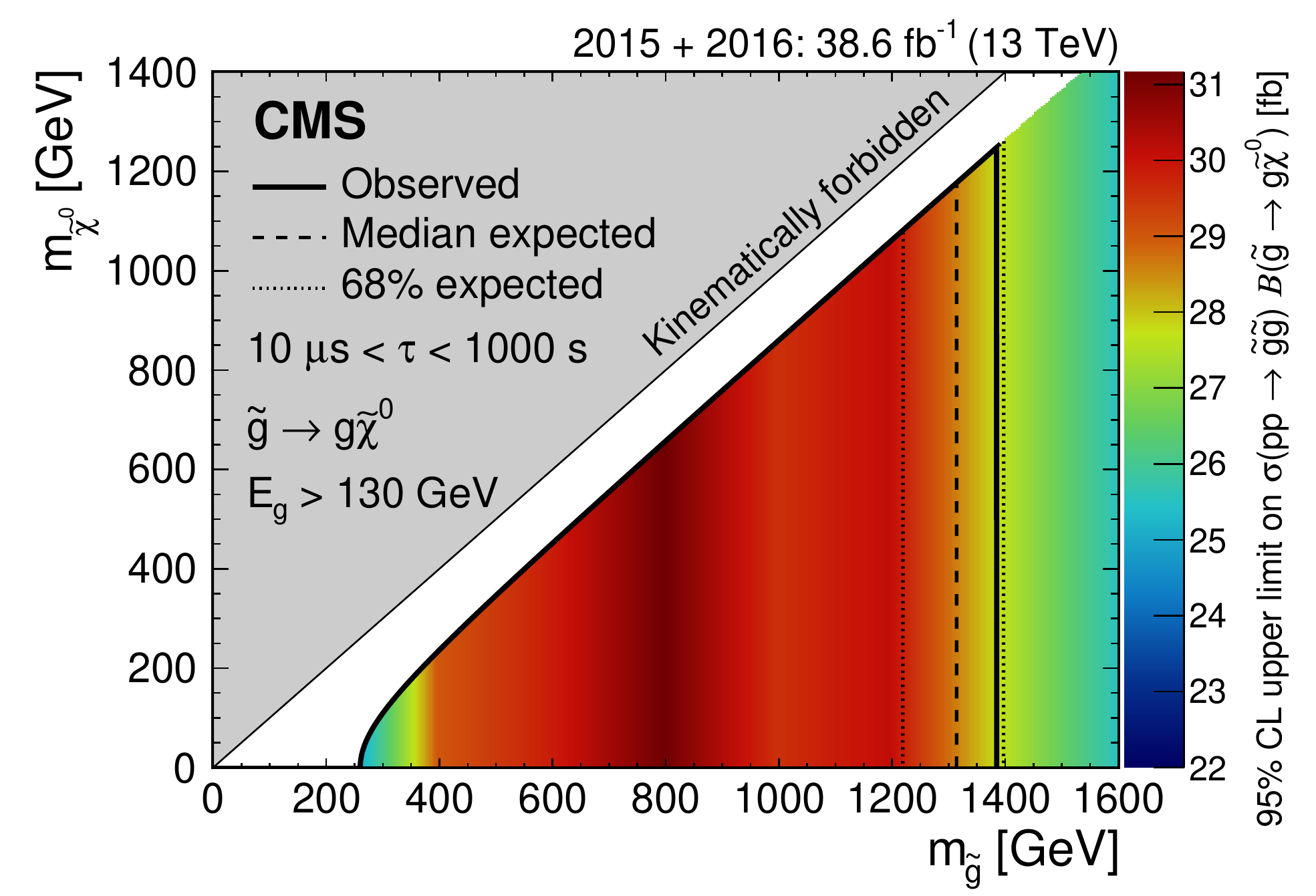}
    \includegraphics[width=0.48\textwidth]{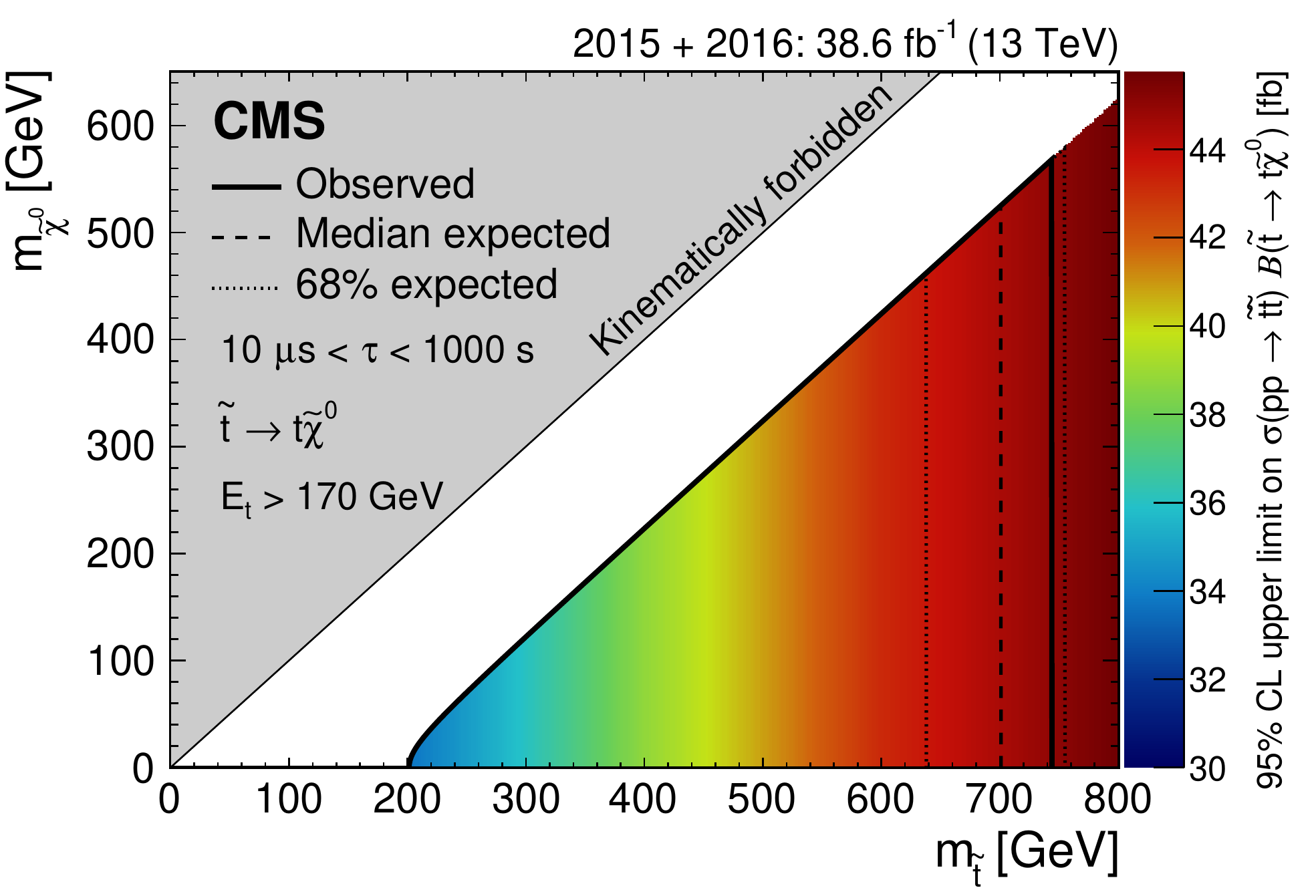}
    \includegraphics[width=0.48\textwidth]{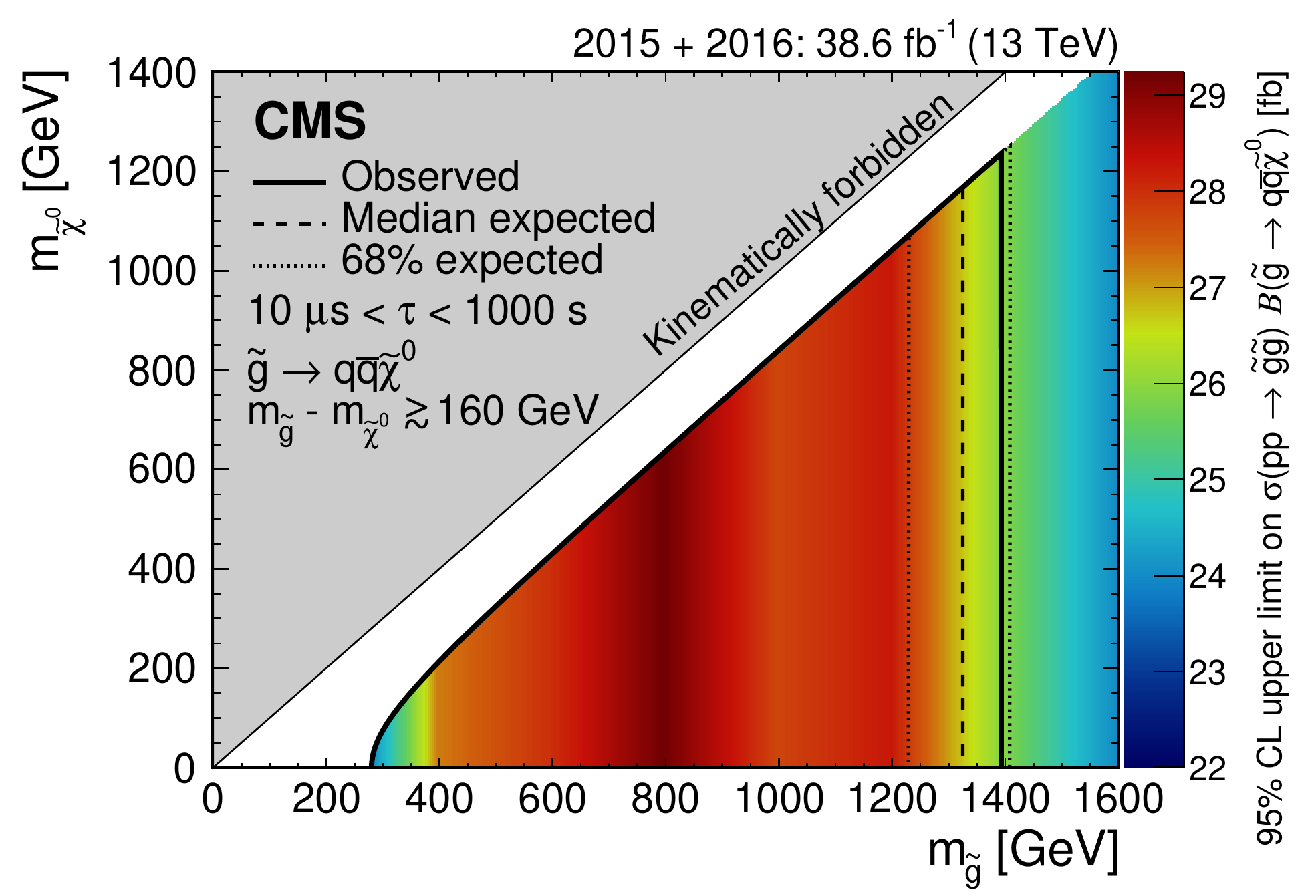}
        \caption{The 95\% CL upper limits in the neutralino mass vs.~gluino (top squark) mass plane, for lifetimes between 10\mus and 1000\unit{s}, for combined 2015 and 2016 data for the calorimeter search. The color map indicates the 95\% CL upper limits on $\mathcal{B}\sigma$. The mostly triangular region defined by the black solid (dashed) line shows the excluded observed (expected) region. 
We show gluinos that undergo a two-body decay (upper \cmsLeft), top squarks that undergo a two-body decay (upper \cmsRight), and gluinos that undergo a three-body decay (lower).}
\label{fig:excludedRegionGluino}
\end{figure}

Figure~\ref{fig:excludedRegionGluino} shows the regions of the gluino (top squark) mass vs.~neutralino mass plane excluded by the calorimeter search, for lifetimes between 10\mus and 1000\unit{s}. The borders of the regions are determined by the edge of the plateau in Fig.~\ref{fig:RecoEff} and the gluino (top squark) mass limits.

\begin{figure}[hbtp]
\centering
\includegraphics[scale=0.39]{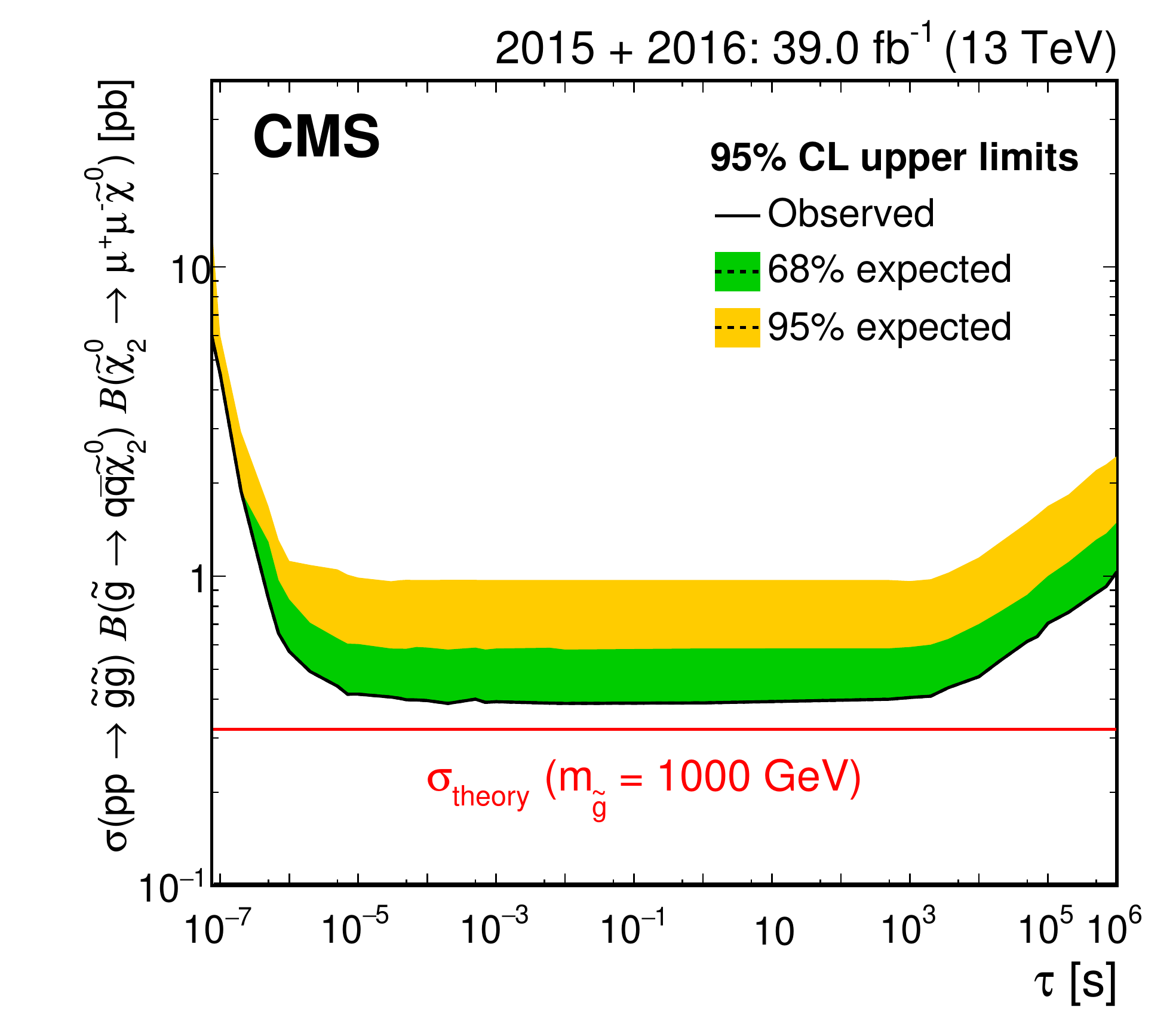}
\includegraphics[scale=0.39]{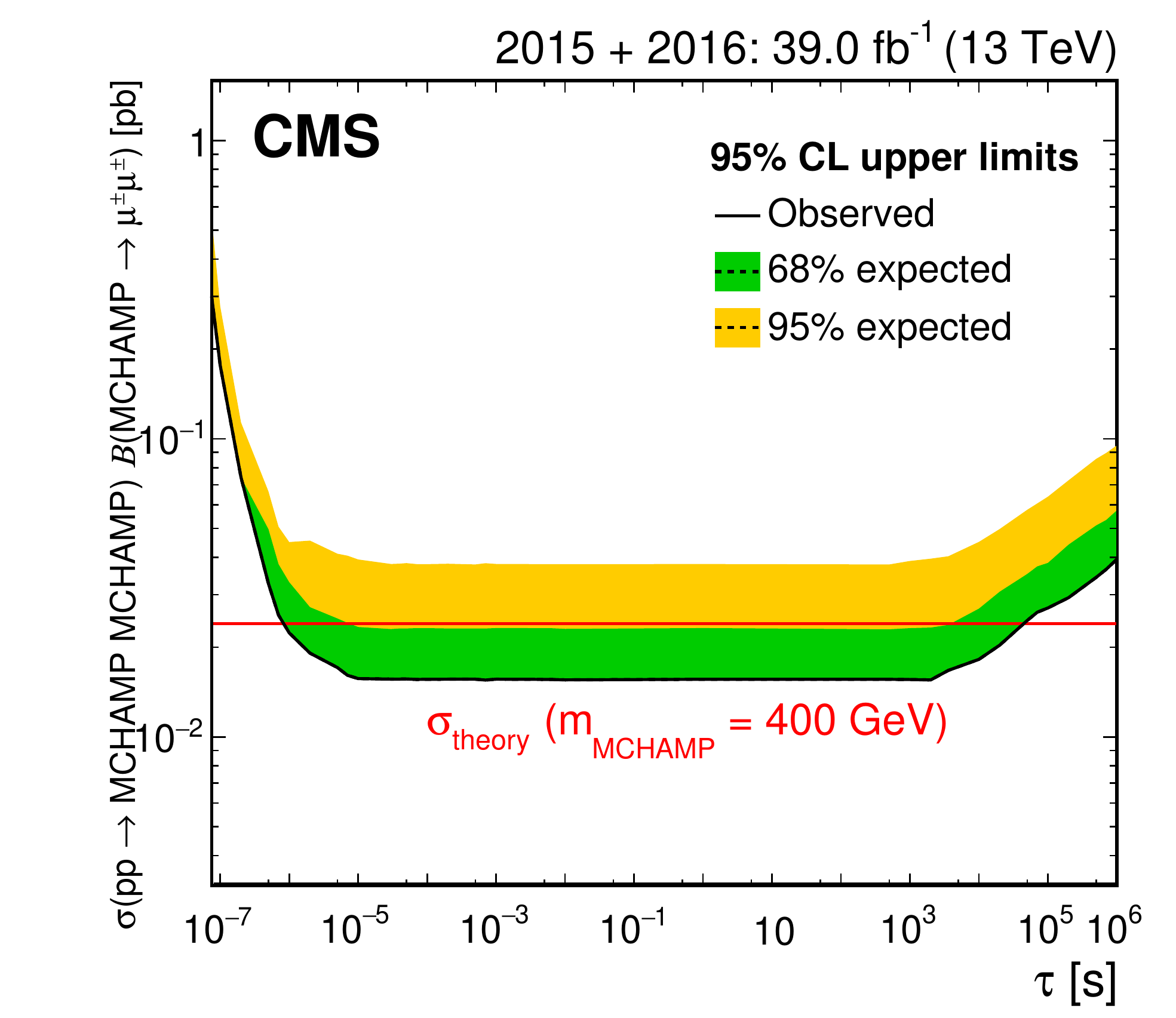}
\caption{\label{fig:xsLifetimeCombined}The 95\% CL upper limits on $\mathcal{B}\sigma$
for 1000\GeV gluino (\cmsLeft) and 400\GeV MCHAMP (\cmsRight) pair production as a function of lifetime,
 for combined 2015 and 2016 data for the muon search. The theory lines assume $\mathcal{B}=100\%$.
}
\end{figure}

\begin{figure}[hbtp]
\centering{}\includegraphics[scale=0.39]{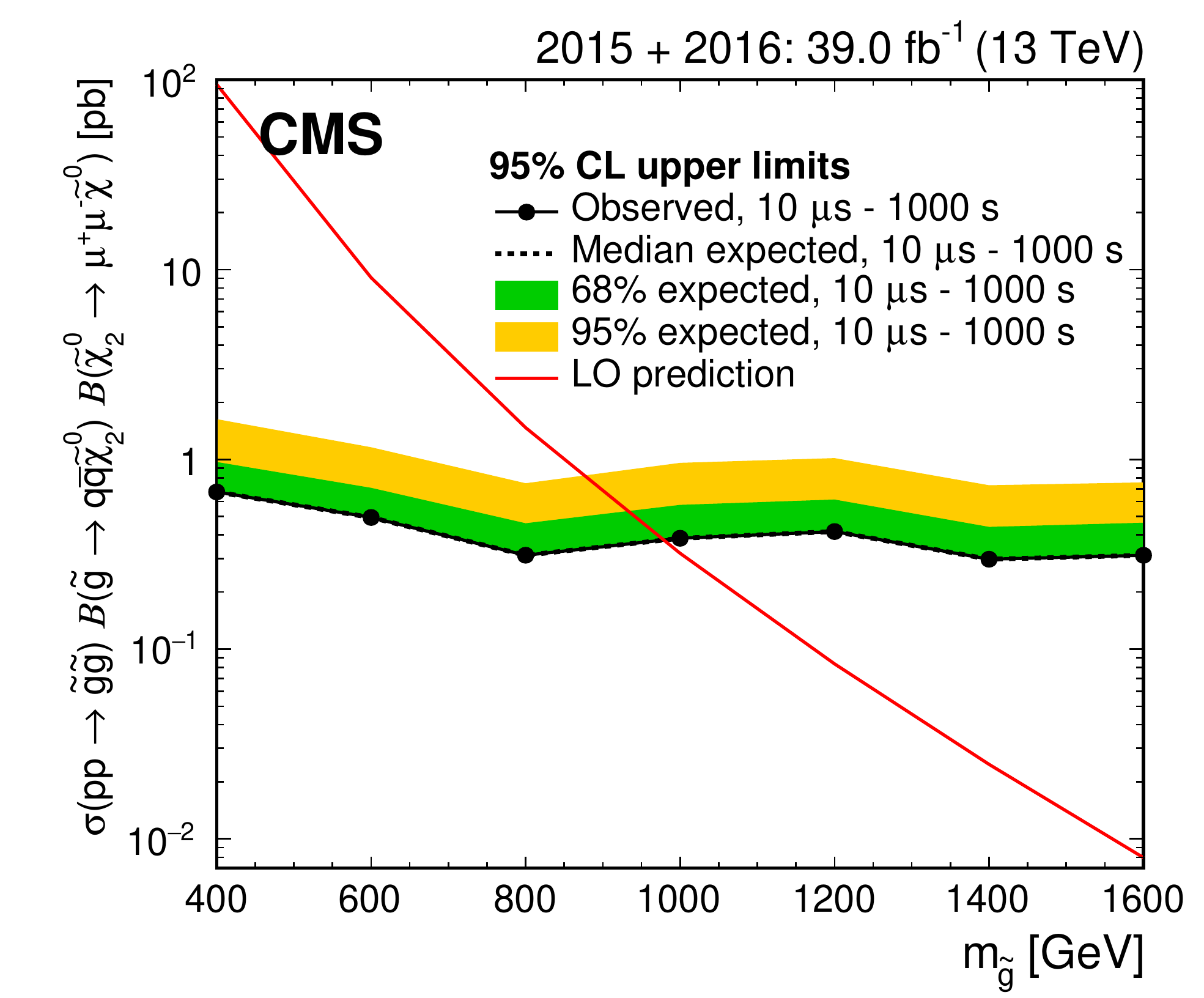}
\centering{}\includegraphics[scale=0.39]{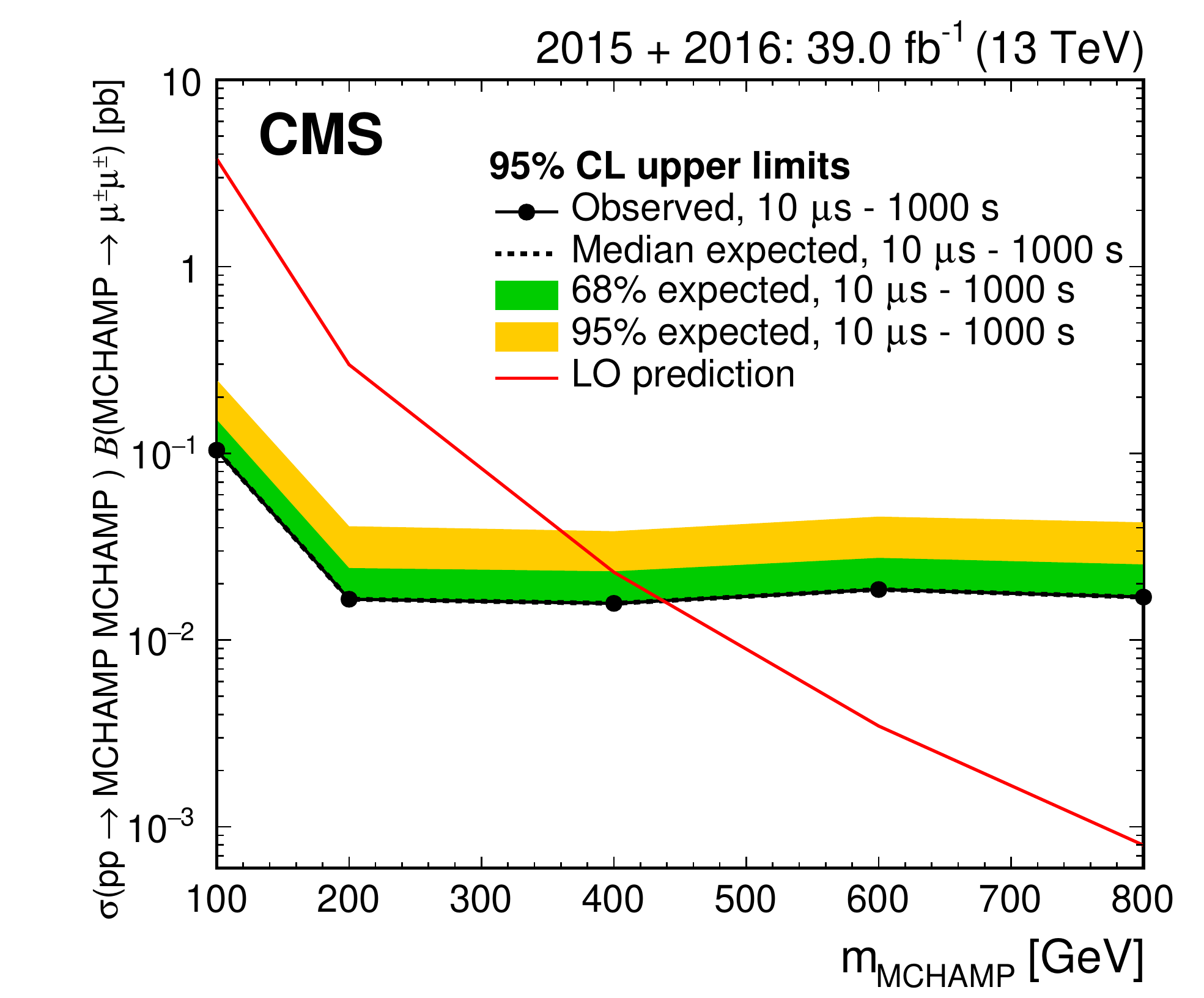}
\caption{\label{fig:xsMassCombination}95\% CL upper limits on $\mathcal{B}\sigma$ for gluino (\cmsLeft) and MCHAMP (\cmsRight) pair production as a function of
mass, for lifetimes between 10\mus and 1000\unit{s}, for combined 2015 and 2016 data for the muon search. The theory curves assume $\mathcal{B}=100\%$.}
\end{figure}

For the muon search, the 95\% CL upper limits on $\mathcal{B}\sigma$ as a function of lifetime
for 1000\GeV gluinos and 400\GeV MCHAMPs are shown in Fig.~\ref{fig:xsLifetimeCombined}
for combined 2015 and 2016 data.
The combined 2015 and 2016 95\% CL upper limits on $\mathcal{B}\sigma$ of gluino and MCHAMP pair production as a function of mass
are shown in Fig.~\ref{fig:xsMassCombination}, for lifetimes between 10\mus and 1000\unit{s}.
Gluinos with masses between 400 and 980\GeV are excluded
for lifetimes between 10\mus and 1000\unit{s}, assuming $\mathcal{B}(\PSg \to \cPq\cPaq\PSGczDt) \mathcal{B}(\PSGczDt \to \MM\PSGcz)=100\%$, $m_{\PSGcz} = 0.25m_{\PSg}$ and $m_{\PSGczDt} = 2.5m_{\PSGcz}$.
MCHAMPs with masses between 100 and 440\GeV and $\abs{Q}=2e$ are excluded
for lifetimes between 10\mus and 1000\unit{s}, assuming $\mathcal{B}(\text{MCHAMP} \to \mu^{\pm}\mu^{\pm})=100\%$.

\section{Summary}
A search has been presented for long-lived particles that stopped in the CMS detector after being produced in proton-proton collisions at a center-of-mass energy of 13\TeV at the CERN LHC. The subsequent decays of these particles to produce calorimeter deposits or muon pairs were looked for during gaps between proton bunches in the LHC beams. In the calorimeter (muon) search, with collision data corresponding to an integrated luminosity of 2.7\,(2.8)\fbinv in a period of sensitivity corresponding to 135\,(155) hours of trigger livetime in 2015 and to an integrated luminosity of 35.9\,(36.2)\fbinv in a period of sensitivity of 586\,(589) hours of trigger livetime in 2016, no excess above the estimated background has been observed. Cross section ($\sigma$) and mass limits have been presented at 95\% confidence level (CL) on gluino (\PSg), top squark (\PSQt), and multiply charged massive particle (MCHAMP) production over 13 orders of magnitude in the mean proper lifetime of the stopped particle.

In the calorimeter search, combining the results from the 2015 and 2016 analyses and assuming a branching fraction ($\mathcal{B}$) of 100\%
for $\PSg \to \cPg\PSGcz$ $(\PSg \to \cPq\cPaq\PSGcz)$, where $\PSGcz$ is the lightest neutralino, gluinos with lifetimes from $10\mus$ to 1000\unit{s} and $m_{\PSg} < 1385$\,(1393)\GeV have been excluded, for a cloud model of R-hadron interactions and for the daughter gluon energy $E_{\cPg} > 130\GeV$ ($m_{\PSg}-m_{\PSGcz} \gtrapprox 160\GeV$). Under similar assumptions, for the daughter top quark energy $E_{\cPqt} > 170\GeV$ and $\mathcal{B}(\PSQt \to \cPqt\PSGcz) = 100\%$, long-lived top squarks with lifetimes from 10\mus to 1000\unit{s} and $m_{\PSQt} < 744\GeV$ have been excluded. These are the first limits on stopped long-lived particles at 13\TeV and the strongest limits to date.

In the muon search,
95\% CL upper limits on $\mathcal{B}\sigma$
were set for combined 2015 and 2016 data.
For lifetimes between 10$\mus$ and 1000\unit{s},
limits were set between 1 and 0.01\unit{pb}
for gluinos with masses between 400 and 1600\GeV and
for MCHAMPs with masses between 100 and 800\GeV
and charge $\abs{Q}=2e$.
For lifetimes between 10$\mus$ and 1000\unit{s},
gluinos with masses between 400 and 980\GeV have been excluded, assuming $\mathcal{B}(\PSg \to \cPq\cPaq\PSGczDt) \mathcal{B}(\PSGczDt \to \MM\PSGcz)=100\%$, $m_{\PSGcz} = 0.25m_{\PSg}$, and $m_{\PSGczDt} = 2.5m_{\PSGcz}$, where $\PSGczDt$ is the next-to-lightest neutralino.
Under the same lifetime hypothesis,
MCHAMPs with masses between 100 and 440\GeV and $\abs{Q}=2e$ have been excluded, assuming $\mathcal{B}(\text{MCHAMP} \to \mu^{\pm}\mu^{\pm})=100\%$.
These are the first limits obtained at the LHC for stopped particles that decay to muons.

\begin{acknowledgments}
We congratulate our colleagues in the CERN accelerator departments for the excellent performance of the LHC and thank the technical and administrative staffs at CERN and at other CMS institutes for their contributions to the success of the CMS effort. In addition, we gratefully acknowledge the computing centers and personnel of the Worldwide LHC Computing Grid for delivering so effectively the computing infrastructure essential to our analyses. Finally, we acknowledge the enduring support for the construction and operation of the LHC and the CMS detector provided by the following funding agencies: BMWFW and FWF (Austria); FNRS and FWO (Belgium); CNPq, CAPES, FAPERJ, and FAPESP (Brazil); MES (Bulgaria); CERN; CAS, MoST, and NSFC (China); COLCIENCIAS (Colombia); MSES and CSF (Croatia); RPF (Cyprus); SENESCYT (Ecuador); MoER, ERC IUT, and ERDF (Estonia); Academy of Finland, MEC, and HIP (Finland); CEA and CNRS/IN2P3 (France); BMBF, DFG, and HGF (Germany); GSRT (Greece); OTKA and NIH (Hungary); DAE and DST (India); IPM (Iran); SFI (Ireland); INFN (Italy); MSIP and NRF (Republic of Korea); LAS (Lithuania); MOE and UM (Malaysia); BUAP, CINVESTAV, CONACYT, LNS, SEP, and UASLP-FAI (Mexico); MBIE (New Zealand); PAEC (Pakistan); MSHE and NSC (Poland); FCT (Portugal); JINR (Dubna); MON, RosAtom, RAS, RFBR and RAEP (Russia); MESTD (Serbia); SEIDI, CPAN, PCTI and FEDER (Spain); Swiss Funding Agencies (Switzerland); MST (Taipei); ThEPCenter, IPST, STAR, and NSTDA (Thailand); TUBITAK and TAEK (Turkey); NASU and SFFR (Ukraine); STFC (United Kingdom); DOE and NSF (USA).

\hyphenation{Rachada-pisek} Individuals have received support from the Marie-Curie program and the European Research Council and Horizon 2020 Grant, contract No. 675440 (European Union); the Leventis Foundation; the A. P. Sloan Foundation; the Alexander von Humboldt Foundation; the Belgian Federal Science Policy Office; the Fonds pour la Formation \`a la Recherche dans l'Industrie et dans l'Agriculture (FRIA-Belgium); the Agentschap voor Innovatie door Wetenschap en Technologie (IWT-Belgium); the Ministry of Education, Youth and Sports (MEYS) of the Czech Republic; the Council of Science and Industrial Research, India; the HOMING PLUS program of the Foundation for Polish Science, cofinanced from European Union, Regional Development Fund, the Mobility Plus program of the Ministry of Science and Higher Education, the National Science Center (Poland), contracts Harmonia 2014/14/M/ST2/00428, Opus 2014/13/B/ST2/02543, 2014/15/B/ST2/03998, and 2015/19/B/ST2/02861, Sonata-bis 2012/07/E/ST2/01406; the National Priorities Research Program by Qatar National Research Fund; the Programa Severo Ochoa del Principado de Asturias; the Thalis and Aristeia programs cofinanced by EU-ESF and the Greek NSRF; the Rachadapisek Sompot Fund for Postdoctoral Fellowship, Chulalongkorn University and the Chulalongkorn Academic into Its 2nd Century Project Advancement Project (Thailand); the Welch Foundation, contract C-1845; and the Weston Havens Foundation (USA).
\end{acknowledgments}

\bibliography{auto_generated}

\cleardoublepage \appendix\section{The CMS Collaboration \label{app:collab}}\begin{sloppypar}\hyphenpenalty=5000\widowpenalty=500\clubpenalty=5000\textbf{Yerevan Physics Institute,  Yerevan,  Armenia}\\*[0pt]
A.M.~Sirunyan, A.~Tumasyan
\vskip\cmsinstskip
\textbf{Institut f\"{u}r Hochenergiephysik,  Wien,  Austria}\\*[0pt]
W.~Adam, F.~Ambrogi, E.~Asilar, T.~Bergauer, J.~Brandstetter, E.~Brondolin, M.~Dragicevic, J.~Er\"{o}, A.~Escalante Del Valle, M.~Flechl, M.~Friedl, R.~Fr\"{u}hwirth\cmsAuthorMark{1}, V.M.~Ghete, J.~Grossmann, J.~Hrubec, M.~Jeitler\cmsAuthorMark{1}, A.~K\"{o}nig, N.~Krammer, I.~Kr\"{a}tschmer, D.~Liko, T.~Madlener, I.~Mikulec, E.~Pree, N.~Rad, H.~Rohringer, J.~Schieck\cmsAuthorMark{1}, R.~Sch\"{o}fbeck, M.~Spanring, D.~Spitzbart, A.~Taurok, W.~Waltenberger, J.~Wittmann, C.-E.~Wulz\cmsAuthorMark{1}, M.~Zarucki
% ...
\vskip\cmsinstskip
% ...
1:~~Also at Vienna University of Technology, Vienna, Austria\\
% ...

\end{sloppypar}
\end{document}